\documentclass[conference]{IEEEtran}
\usepackage{fancyhdr}
\usepackage[table,dvipsnames]{xcolor}
\usepackage{svg}
\usepackage{algorithm}
\usepackage{algpseudocode}
\algrenewcommand{\algorithmiccomment}[1]{\hfill\(\triangleright\)~\textit{#1}}
\usepackage{bm}
\usepackage{booktabs,tabularx}
\usepackage{capt-of}
\usepackage{multirow}
\usepackage{float}
\usepackage{siunitx}
\usepackage{subcaption}
\usepackage{amsmath}
\usepackage{tikz}
\usetikzlibrary{arrows.meta,positioning,fit,shapes.geometric,calc} 
\usepackage{adjustbox}
\usepackage{lipsum,soul}
\usepackage{amsfonts}
\usepackage{graphicx}
\usepackage{epstopdf}
\usepackage{anyfontsize}
\usepackage{comment}  
\usepackage{amsthm}
\usepackage{amssymb}
\usepackage[noadjust]{cite}
\usepackage{url}
\usepackage{hyperref}
\usepackage{cleveref}
\usepackage{balance}
\usepackage{xurl}

\algrenewcommand\algorithmicrequire{\textbf{Input:}}
\algrenewcommand\algorithmicensure{\textbf{Output:}}

\newcolumntype{H}{@{}>{\setbox0=\hbox\bgroup}c<{\egroup}@{}}

\theoremstyle{plain}

\newtheorem{lemma}{Lemma}

\newtheorem{definition}{Definition}

\newcommand{\bb}{\bm{b}}

\AtBeginDocument{%
  }

\begin{document}

\title{Tackling Parallelization Challenges of Randomized Preconditioners with Dependency Tracking}

\makeatletter
\newcommand{\linebreakand}{%
  \end{@IEEEauthorhalign}
  \hfill\mbox{}\par
  \mbox{}\hfill\begin{@IEEEauthorhalign}
}
\makeatother

\author{
  \IEEEauthorblockN{Tianyu Liang}
  %\thanks{$^{*}$ Lawrence Berkeley National Lab}
  \IEEEauthorblockA{University of California, Berkeley\\
  Berkeley, California, USA\\
  tianyul@berkeley.edu}
  \and
  \IEEEauthorblockN{Chao Chen}
  \IEEEauthorblockA{North Carolina State University\\
  Raleigh, North Carolina, USA\\
  cchen49@ncsu.edu}
  \and
  \IEEEauthorblockN{Yotam Yaniv}
  \IEEEauthorblockA{Lawrence Berkeley National Lab\\
  Berkeley, CA, USA\\
  yotamy@lbl.gov}
  \linebreakand
  \IEEEauthorblockN{Hengrui Luo}
  \IEEEauthorblockA{Rice University\\
  Houston, Texas, USA\\
  hl180@rice.edu}
  \and
  \IEEEauthorblockN{David Tench}
  \IEEEauthorblockA{Lawrence Berkeley National Lab\\
  Berkeley, California, USA\\
  dtench@lbl.gov}
  \and
  \IEEEauthorblockN{Xiaoye S. Li}
  \IEEEauthorblockA{Lawrence Berkeley National Lab\\
  Berkeley, California, USA\\
  XSLi@lbl.gov}
  \linebreakand
  \IEEEauthorblockN{Ayd{\i}n Bulu\c{c}}
  \IEEEauthorblockA{Lawrence Berkeley National Lab\\
  Berkeley, California, USA\\
  abuluc@lbl.gov}
  \and
  \IEEEauthorblockN{James Demmel}
  \IEEEauthorblockA{University of California, Berkeley\\
  Berkeley, California, USA\\
  demmel@berkeley.edu}
  \and
  \IEEEauthorblockN{Jie Chen$^{\dagger}$
  \thanks{$^{\dagger}$ Work was done when affiliated with MIT-IBM Watson AI Lab, IBM Research.}}
  \IEEEauthorblockA{Amazon Web Services\\
  Seattle, Washington, USA\\
  jiechenj@amazon.com}
}

\IEEEoverridecommandlockouts
\maketitle
\thispagestyle{fancy}
\lhead{}
\rhead{}
\chead{}
\lfoot{\footnotesize{
SC26, November 15-20, 2026, Chicago, Illinois, USA
\newline 979-8-3195-4789-7/26/\$31.00 \copyright 2026 IEEE}}
\rfoot{}
\cfoot{}
\renewcommand{\headrulewidth}{0pt}
\renewcommand{\footrulewidth}{0pt}

\begin{abstract}
%Solving large, sparse linear systems with respect to the graph Laplacian matrix has important applications in many fields, including physical sciences and engineering. When an iterative solver is used, critical to its rapid convergence is the preconditioner that approximates the (pseudo)inverse of the system matrix. Recently, randomized preconditioners based on incomplete factorizations were proposed and theoretically studied. However, their high-performance implementation is challenging because fill-ins are stochastic and unpredictable, unfriendly to parallelization. In this work, we develop an efficient parallelization approach based on dynamic dependency tracking, amenable to both CPU and GPU architectures. We devise and demonstrate that our approach, named \texttt{ParAC}, achieves a higher degree of parallelism than existing parallel implementations of non-randomized preconditioners compared to the standard incomplete Cholesky factorization. Furthermore, \texttt{ParAC} converges significantly faster than incomplete Cholesky implemented in optimized software, including MATLAB (via CPU) and \texttt{cuSPARSE} (via GPU), in terms of the running time.
Incomplete Cholesky (IC) is a common memory-efficient preconditioner for large graph Laplacian systems, but classical IC can break down on irregular graphs. Approximate Cholesky (AC) addresses this with randomized dropping that is provably breakdown-free. Its stochastic fill-in pattern, however, resists static parallelization, creating a ``setup bottleneck.'' We present \texttt{ParAC}, which parallelizes the AC sampling rule of Gao, Kyng, and Spielman using dynamic dependency tracking and architecture-aware memory management to execute it efficiently on CPUs and GPUs while preserving its numerical updates. Total time-to-solution speedups over deterministic IC reach \textbf{57.3$\times$} on CPU and \textbf{52.2$\times$} on GPU. Algebraic multigrid (AMG) generally remains preferable on structured PDE discretizations. On the irregular graph Laplacians tested, \texttt{ParAC} is up to \textbf{7.2$\times$} faster than AMG on CPUs and \textbf{6.0$\times$} faster on GPUs. On 200 heterogeneous SDDM systems, \texttt{ParAC} converges on every instance, whereas the AMG solvers fail on 15 and 19, on CPU and GPU, respectively.

\end{abstract}

%--------------------------------------------------------------------
\section{Introduction}
\label{sec:problem_def}

Large-scale, sparse symmetric diagonally dominant (SDD) systems arise in spectral graph partitioning~\cite{spectral-partition}, graph-based machine learning~\cite{JMLR:learn-lap, learn-graph, coarsening}, and numerical solutions of partial differential equations (PDEs)~\cite{elliptic-near-linear}. In this paper, we focus on solving the linear system
\begin{equation} \label{eq:lxb}
L x = b,
\end{equation}
where $L$ is a \textbf{graph Laplacian}. General SDD systems can be reduced to Laplacian systems with constant-factor overhead~\cite{gremban1996combinatorial}.

\textbf{Solver Landscape: Direct vs. Iterative.}
State-of-the-art methods for solving \eqref{eq:lxb} generally fall into two categories: (sparse) direct and iterative solvers. Sparse direct methods~\cite{Davis2016DirectSurvey, george1981computer}, such as Cholesky factorization, are valued for their numerical reliability. However, they frequently encounter a ``memory wall'' due to \textbf{fill-in}, that is, the generation of new non-zero entries during Gaussian elimination. This leads to prohibitive computational and spatial overheads for solving large-scale systems.
Iterative methods offer a memory-efficient alternative but require high-quality preconditioners to achieve competitive convergence rates~\cite{saad-iterative}. For graph Laplacians (and more generally, symmetric positive semi-definite matrices), a general, algebraic preconditioning approach is Incomplete Cholesky (IC) factorization~\cite{incomplete-chol}, which constructs an approximation
\begin{equation} \label{eq:ic}
L \approx G G^\top,
\end{equation}
where $G$ is a lower triangular matrix. Traditional IC drops fill-ins during elimination based on either a pre-determined sparsity pattern (level-based)~\cite{Watts1981} or a prescribed numerical threshold~\cite{Munksgaard1980, JonesPlassmann1995}. While simple to implement, deterministic IC can be numerically unstable; dropping entries may destroy the positive semi-definiteness of intermediate Schur complements, causing the factorization to break down~\cite{manteuffel1980incomplete}.

\textbf{Breakthrough of Approximate Cholesky (AC).}
Recent breakthroughs have shown that IC using \textbf{randomized dropping rules} produces significantly more robust preconditioners. We refer to this family of methods collectively as \textbf{Approximate Cholesky (AC)}~\cite{kyng2016approximate, gao2023robust, rchol}. The AC variants considered here are unbiased, satisfying $\mathbb{E}[G G^\top] = L$. Unlike deterministic IC, they are also provably \textbf{breakdown-free} on any connected graph Laplacian~\cite{rchol, gao2023robust}. The reason is that the randomized update always retains a spanning tree over the eliminated vertex's neighborhood. Empirically, AC also converges significantly faster than deterministic dropping rules across a wide variety of problems~\cite{rchol, gao2023robust}. \texttt{ParAC} parallelizes the AC sampling rule of Gao, Kyng, and Spielman~\cite{gao2023robust}, which is also used by \texttt{rchol} (randomized Cholesky factorization)~\cite{rchol}; \Cref{sec:randomized} reviews the method and its provenance in detail.

\textbf{Parallelism Gap and Setup Bottleneck.} Despite its numerical superiority, AC faces a significant hurdle on modern high-performance architectures, particularly GPUs. In the preconditioned iterative phase, triangular solves are now highly optimized and iteration counts are kept low with a high-quality AC preconditioner. Consequently, \textbf{the factorization (setup) phase frequently emerges as the primary computational bottleneck}. If the factorization remains serial or poorly parallelized, it negates the speedup of the subsequent solve phase.

To harness the full potential of randomized preconditioning, we need parallel algorithms that maintain AC’s numerical robustness while achieving high performance. However, AC's stochastic nature makes this difficult: unlike level-based IC, where fill-in positions are known a priori, AC’s fill-in pattern is determined ``online,'' so a static schedule cannot represent the realized dependencies exactly.

\subsection{Existing Parallel Methods and Limitations}

A substantial body of work parallelizes fixed-pattern or level-scheduled incomplete Cholesky and incomplete LU factorizations on CPUs~\cite{HysomPothen1999,HysomPothen2001,Haase1998,DoiWashio1999,intel_wavefront} and GPUs~\cite{naumov2012parallelLU}. These methods exploit dependencies that are known before numerical factorization. Other GPU work uses two-level or structure-aware ILU schemes~\cite{XuLiOseiKuffuor2025,luo2025structilu}.

Chow and Patel~\cite{chow2015fine} introduced fine-grained fixed-point sweeps over a prescribed factor pattern. ParILUT later added adaptive pattern thresholding, with ParICT as its symmetric Cholesky variant~\cite{anzt2018parilut}; ParILUT has also been implemented on GPUs~\cite{AnztRibizelFlegarChowDongarra2019}. These methods differ from the conventional elimination-time threshold dropping used by our MATLAB \texttt{ichol} baseline. Their fixed-point-sweep execution model, including ParILUT's adaptive pattern refinement, updates factor entries rather than sampling clique updates on an evolving active graph, so it does not transfer directly to AC.

Domain decomposition is another strategy for parallel incomplete factorization, especially in distributed-memory settings~\cite{MagoluVanDerVorst2001FGCS,MagoluVanDerVorst2001ParCo}. The closest prior work, \texttt{rchol}~\cite{rchol}, applies this idea to AC on multi-core CPUs: a nested-dissection ordering partitions the graph so that subdomains are eliminated concurrently while the separators are eliminated sequentially. Such static schemes share an inherent drawback: partitioning carries a nontrivial preprocessing cost. \texttt{rchol} adds two more. The separator work remains serial, and the scheme does not map onto the fine-grained parallelism a GPU requires. These costs are especially pronounced when separators are imbalanced or the graph admits no useful separator hierarchy. \texttt{ParAC} instead tracks dependencies dynamically, which removes the partitioning step entirely, eliminates the serial separator phase, and admits an efficient GPU implementation.

Related theoretical work by Sachdeva and Zhao parallelizes the Kyng--Sachdeva solver through block Cholesky elimination and random sampling~\cite{sachdeva2023simple}; this construction differs from the practical one-vertex-at-a-time AC kernel of Gao, Kyng, and Spielman~\cite{gao2023robust} considered here. Baumann and Kyng~\cite{baumann2024} analyzed the critical-path depth of a parallel AC algorithm that samples low-degree vertices and eliminates an independent subset in each step. Although this approach is theoretically appealing, translating it into an efficient parallel implementation remains challenging.

Dynamic data structures and runtime scheduling already appear in sparse factorization~\cite{superlu-overview}. Unlike fixed-pattern methods, however, AC cannot precompute an exact schedule for the fill that will actually be sampled. \texttt{ParAC}'s contribution is the AC-specific readiness invariant of \Cref{parallel:insight} and its realization on CPUs and GPUs.

\subsection{Contributions: \texttt{ParAC}} 

In this paper, we introduce \texttt{ParAC}, a practical high-performance execution framework built on \textbf{dynamic dependency tracking} and \textbf{architecture-aware memory management}. For a prescribed elimination order, \texttt{ParAC} preserves the numerical updates and output distribution of the AC sampling rule of Gao, Kyng, and Spielman~\cite{gao2023robust}, and therefore inherits its breakdown-free guarantee~\cite{gao2023robust,rchol}. At the same time, it exposes finer-grained parallelism than \texttt{rchol}'s nested-dissection scheme and maps the AC sampling rule to GPUs. Our aim is not to displace algebraic multigrid (AMG). IC and AMG are complementary tools~\cite{benzi2002preconditioning}. On suitable structured PDE families, AMG can exploit smooth error components to achieve mesh-independent convergence~\cite{RugeStuben1987,HensonYang2002}, while its performance and memory use can be less predictable on irregular graphs. \texttt{ParAC} offers a lightweight, memory-efficient alternative for graph Laplacians and SDDM systems without a numerical drop tolerance. Our contributions are as follows:
\begin{itemize}
\item \textbf{The \texttt{ParAC} Framework:} We introduce a dynamic dependency-tracking execution model that identifies the direct dependencies of each vertex (column) at runtime. This allows \texttt{ParAC} to schedule elimination tasks with low overhead, avoiding the conservative bottlenecks of static elimination trees~\cite{liu1990role,Davis2016DirectSurvey}, whose exact-fill dependencies are too restrictive for efficient AC scheduling.
\item \textbf{Architecture-Aware Design:} We realize this model on both multicore CPUs and massively parallel GPUs, using lock-free dependency updates and pre-allocated memory structures to avoid the severe penalties of dynamic allocation inside parallel CUDA kernels. On the GPU we further adopt a persistent-kernel model, computing the entire factorization within a single launch so that kernel launch overhead remains negligible even for large systems.
\item \textbf{Empirical Evaluation:} We evaluate \texttt{ParAC} on the NERSC Perlmutter supercomputer against deterministic IC and AMG solvers~\cite{HensonYang2002,RugeStuben1987}. An ordering study shows that fill-reducing AMD~\cite{amd-order} is a strong CPU default, whereas NNZ-sorted ordering consistently wins in our GPU tests. Against deterministic IC, \texttt{ParAC} attains end-to-end speedups of up to \textbf{57.3$\times$} on CPU and \textbf{52.2$\times$} on GPU, measured where the baseline itself converges. On the irregular graph Laplacians tested, it is up to \textbf{7.2$\times$} faster than AMG on CPUs and \textbf{6.0$\times$} faster on GPUs. It also remains robust on massive real-world graphs where classical IC fails to converge and AMG exhausts GPU memory.
\end{itemize}

Our code is available at \url{https://github.com/Tianyu-Liang/Parallel-Randomized-Cholesky}.

%--------------------------------------------------------------------
\section{Background}

In this section, we establish the foundations of the Approximate Cholesky (AC) algorithm. We focus on the relationship between the algebraic elimination steps and their corresponding graph-theoretic interpretations, which dictate the data dependencies explored in this work.

\subsection{Graph Laplacians and Graph-Matrix Duality}
We begin with the graph Laplacian, defined as follows:

\begin{definition}[Graph Laplacian]
A matrix $L \in \mathbb{R}^{N \times N}$ is a graph Laplacian if (1) $L$ is symmetric, (2) the row sums are zero ($\sum_{j} \ell_{ij} = 0$), and (3) all off-diagonal entries are non-positive ($\ell_{ij} \le 0$ for $i \neq j$).
\end{definition}

There is a one-to-one mapping between a Laplacian $L$ and an undirected, weighted graph $\mathcal{G} \triangleq (V, E, w)$. Each row/column index corresponds to a vertex, and each non-zero off-diagonal entry $\ell_{ij}$ represents an edge between vertices $i$ and $j$ with weight $w_{ij} \triangleq -\ell_{ij}$. Alternatively, $L$ can be expressed as a sum of outer products representing individual edges:
\begin{equation}
L 
= \sum_{(i,j) \in E} w_{ij} (\mathbf{e}_i 
- \mathbf{e}_j)(\mathbf{e}_i - \mathbf{e}_j)^\top,
\end{equation}
where $\mathbf{e}_i$ is the $i$-th standard basis vector in $\mathbb{R}^{N}$.

For the remainder of this paper, we assume $\mathcal{G}$ is connected (i.e., $L$ is irreducible). Under this assumption, $L$ has a single zero eigenvalue with a null space spanned by the all-ones vector. As shown in~\cite{rchol}, AC is robust to this singularity; pivots only vanish at the final step, and Krylov solvers remain stable provided the system $Lx=b$ is consistent ($b \perp \mathbf{1}$).

% Such cliques lead to significant computational and memory overhead. As the factorization proceeds step by step, cliques are expanded with more and more vertices. For large matrices, the introduced fill-ins significantly slow down the factorization~\cite{kyng2016approximate, davis2006direct, george1981computer, liu1990role}. What is worse, in parallel processing, the accumulation of fill-ins becomes a memory bottleneck and limits parallelization opportunities~\cite{davis2006direct, liu1990role}. This phenomenon motivates the development of incomplete factorizations, which drop fill-ins. In this work, we are concerned with randomly dropping the fill-ins.

%--------------------------------------------------------------------
\subsection{From Clique Formation to Clique Sampling}
\label{sec:randomized}

Consider applying one step of Cholesky factorization (Gaussian elimination) to a Laplacian matrix on row/column $k$.
Algebraically, this forms a Schur complement. Graphically, the neighbors of $k$ form a \textbf{clique}~\cite{davis2006direct, liu1990role}, that is every neighbor is now connected to every other neighbor.
This ``clique formation'' is the source of \textbf{fill-in}. In standard Cholesky, a vertex with $d$ neighbors generates $O(d^2)$ new edges. In large-scale systems, these cliques quickly become dense, creating a ``memory wall'' and complex dependency chains that serialize the factorization process.

AC, summarized in Algorithms~\ref{a:rchol} and \ref{a:sample}, belongs to the IC family: it computes a factorization of the form~\eqref{eq:ic}, and differs from classical IC in how it sparsifies each clique update. In place of the $O(d^2)$ clique update it inserts a stochastically sampled spanning tree of only $O(d)$ edges.

\paragraph{Provenance}
Kyng and Sachdeva introduced approximate Cholesky with a spectral-approximation guarantee~\cite{kyng2016approximate}. The practical AC sampling rule used here is due to Gao, Kyng, and Spielman~\cite{gao2023robust}, building on that framework. The \texttt{rchol} solver implements this rule and credits its development to Spielman and Kyng (private communication, 2020)~\cite{rchol}; it also provides a nested-dissection parallelization on multi-core CPUs. \texttt{ParAC} parallelizes this AC sampling rule without changing its numerical updates.

\begin{algorithm}[!t]
\caption{Approximate Cholesky~\cite{gao2023robust}, based on~\cite{kyng2016approximate}}
\label{a:rchol}
\begin{algorithmic}[1]
\Require Graph Laplacian matrix $L \in \mathbb{R}^{N \times N}$
\Ensure Lower triangular factor $G \in \mathbb{R}^{N \times N}$ ($L \approx GG^{\top}$)
\State $G = \bm{0}_{N \times N}$
\For {$k=1$ \textbf{to} $N-1$}
\Comment{$N-1$ iterations}
\State $G(:,k) = L(:,k)/ \sqrt{\ell_{kk}}$
%\State{$D(k, k) = \ell_{kk}$}
\Comment{$\ell_{kk}>0$~\cite{rchol}}
\State $L = L - L^{(k)} + \texttt{SampleClique}(L, k)$ \label{line:eliminate}
\Comment{see~\eqref{e:lk}}
\EndFor
\end{algorithmic}
\end{algorithm}

\begin{algorithm}[!t]
\caption{\texttt{SampleClique}: Approximate Cholesky~\cite{gao2023robust}, based on~\cite{kyng2016approximate}}
\label{a:sample}
\begin{algorithmic}[1]
\Require Graph Laplacian matrix $L \in \mathbb{R}^{N \times N}$, index $k$
\Ensure Graph Laplacian  $C \in \mathbb{R}^{N \times N}$  for sampled edges
\State $C = \bm{0}_{N \times N}$
\State $\mathcal{N}_k = \{i \mid i \neq k,\ \ell_{ki} \neq 0\}$ \Comment{neighbors of $k$}
\State Sort $\mathcal{N}_k$ in ascending order based on $|\ell_{ki}|$ for $i \in \mathcal{N}_k$ \label{line:sort}
\State $S=\ell_{kk}$
%\Comment{\cyan{// $\ell_{kk} = - \sum_{i \in \mathcal{N}_k} \ell_{ki}$}}
\While{$ \left| \mathcal{N}_k \right| > 1$}
\Comment{loop over neighbors}
\State Let $i$ be the first element in $\mathcal{N}_k$
\State{$\mathcal{N}_k = \mathcal{N}_k \backslash \{i\}$} 
\Comment{remove $i$ from the set}
\State $S=S+\ell_{ki}$
%\Comment{\cyan{// $S = - \sum_{j \in \mathcal{N}_k} \ell_{kj}$}}
\State{Sample  $j$ from $\mathcal{N}_k$ with {probability $|\ell_{kj}|/S$}}
\State $C = C - \frac{S\, \ell_{ki}}{\ell_{kk}} \, \bb_{ij} \bb_{ij}^\top$
\Comment{select weighted edge $(i,j)$}
\EndWhile
\end{algorithmic}
\end{algorithm}

The critical challenge for parallelism lies in Line~\ref{line:eliminate} of Algorithm~\ref{a:rchol}: the edges drawn by \texttt{SampleClique} are \textbf{stochastic}, so the dependency structure is generated online (\Cref{parallel:insight}).

To formalize the mechanics of AC, we first define the local Laplacian $L^{(k)}$, which represents the star graph centered at vertex $k$ and its neighbors $\mathcal{N}_k \triangleq \{i \mid i \neq k,\ \ell_{ki} \neq 0\}$:
\begin{equation} \label{e:lk}
L^{(k)} \triangleq \sum_{i \in \mathcal{N}_k} \, (-\ell_{ki}) \,\, \bb_{ki} \bb_{ki}^\top,
\end{equation}
where $\mathbf{b}_{ki} \triangleq \mathbf{e}_k - \mathbf{e}_i$ denotes the difference between two standard basis vectors. The symmetry of the outer product $(\mathbf{e}_k - \mathbf{e}_i)(\mathbf{e}_k - \mathbf{e}_i)^\top$ ensures that the edge orientation does not affect the resulting Laplacian.

The corresponding exact clique update is $C_{\mathrm{exact}}^{(k)} \triangleq L^{(k)}-\ell_{kk}^{-1}L(:,k)L(k,:)$.

Replacing the full clique with $O(|\mathcal{N}_k|)$ sampled edges reduces the update size for each pivot. The total work of the practical algorithm still depends on the degrees generated by the prescribed elimination order. The sampling is not arbitrary: the $|\mathcal{N}_k| - 1$ edges drawn by \Cref{a:sample} always form a \textbf{spanning tree} over the neighborhood $\mathcal{N}_k$, and their weights are chosen so that the sampled update is itself a graph Laplacian. Because the active graph therefore stays a connected Laplacian at every step, the factorization is breakdown-free~\cite[Thm.~2.7, Cor.~2.8]{rchol}, \cite[Claim~3.5, Thm.~3.6]{gao2023robust}.

\paragraph{Theoretical Qualification} For a prescribed elimination order, \texttt{ParAC} changes only the scheduling of the AC algorithm in Algorithms~\ref{a:rchol} and \ref{a:sample}, preserving its sampling distribution and breakdown-free guarantee~\cite{gao2023robust,rchol}. Conditioned on the current active graph, each sampled update has expectation $C_{\mathrm{exact}}^{(k)}$, making the resulting factorization unbiased: $\mathbb{E}[GG^\top]=L$. This expectation identity controls the mean but not how far an individual sampled factor may deviate. The spectral analysis of Kyng and Sachdeva additionally relies on multiedge splitting and a uniformly random elimination order. The single-tree AC variant used here and in \texttt{rchol} omits the former, and the configurations evaluated here use AMD or NNZ-sorted order rather than the latter, so that analysis does not apply directly~\cite{kyng2016approximate,gao2023robust,rchol}.

Empirically, the PCG iteration counts reported for \texttt{rchol} on refined three-dimensional Poisson problems grow approximately logarithmically with $N$~\cite[Tabs.~7--8]{rchol}. For a fixed stopping tolerance, this behavior is consistent with---but does not establish---$O(\log^2 N)$ growth in the preconditioned condition number under standard worst-case PCG convergence estimates. We therefore make no corresponding spectral-guarantee claim and evaluate preconditioner quality empirically.

\section{Dynamic Dependency Tracking}
\label{parallel:insight}

Dynamic dependency tracking is the key mechanism to expose parallelism despite AC's stochastic fill-ins. Instead of relying on a static symbolic schedule, \texttt{ParAC} tracks the active dependency structure online and schedules a vertex as soon as its  dependencies vanish.

% \begin{figure}[t]
%   \centering
%   \includegraphics[height=0.7\linewidth,width=1.0\linewidth]{fig/example_graph.png}
%   \caption{Top left: adjacency graph representation of an input matrix. Top right: the corresponding elimination tree. Bottom: a possible graph after eliminating vertex $0$ using AC. Instead of forming a clique among the neighbors of $0$, AC forms a sampled spanning tree (highlighted in red).}
%   \label{fig:mat_pattern}
% \end{figure}

\begin{figure}[t] 
\centering
\resizebox{0.97\columnwidth}{!}{%
\begin{tikzpicture}[
    x=1cm,y=1cm,
    >=Latex,
    every node/.style={font=\small},
    vblue/.style={circle, fill=cyan!35!blue!25, draw=none, minimum size=6.5mm, inner sep=0pt},
    vgreen/.style={circle, fill=green!45!blue!20, draw=none, minimum size=6.5mm, inner sep=0pt},
    edgegray/.style={draw=gray!70, line width=1.1pt},
    edgered/.style={draw=red!85!black, line width=1.5pt},
    treeedge/.style={-Latex, draw=black, line width=0.95pt},
    box/.style={draw=gray!50, rounded corners=2pt, inner sep=3pt, align=left}
]

% =========================
% (a) elimination tree
% =========================
\begin{scope}[shift={(1,0)}]
\node[anchor=west,font=\bfseries] at (-0.3,2.3) {(a) Elimination tree};

\node[vgreen] (b0) at (0,1.7) {0};
\node[vgreen] (b1) at (0,0.85) {1};
\node[vgreen] (b3) at (0,0.0) {3};
\node[vgreen] (b4) at (0,-0.85) {4};
\node[vgreen] (b5) at (1.7,0.0) {5};
\node[vgreen] (b2) at (1.7,1.7) {2};

\draw[treeedge] (b0)--(b1);
\draw[treeedge] (b1)--(b3);
\draw[treeedge] (b3)--(b4);
\draw[treeedge] (b4)--(b5);
\draw[treeedge] (b2)--(b5);
\end{scope}

% =========================
% (b) input graph
% =========================
\begin{scope}[shift={(6.0,0)}]
\node[anchor=west,font=\bfseries] at (-1.5,2.3) {(b) Input graph};

\node[vblue] (a4) at (-1.3,0.2) {4};
\node[vblue] (a0) at (0,0) {0};
\node[vblue] (a3) at (-0.7,-0.95) {3};
\node[vblue] (a1) at (1.2,-0.35) {1};
\node[vblue] (a5) at (0.7,0.95) {5};
\node[vblue] (a2) at (1.55,1.7) {2};

\draw[edgegray] (a0)--(a4);
\draw[edgegray] (a0)--(a3);
\draw[edgegray] (a0)--(a1);
\draw[edgegray] (a0)--(a5);
\draw[edgegray] (a5)--(a2);
\end{scope}

% =========================
% (c) after eliminating 0
% =========================
\begin{scope}[shift={(10.0,0)}]
\node[anchor=west,font=\bfseries] at (-0.7,2.3) {(c) After eliminating $0$};

\node[vblue] (c1) at (0,1.5) {1};
\node[vblue] (c5) at (0,0.35) {5};
\node[vblue] (c2) at (0,-0.95) {2};
\node[vblue] (c3) at (-1.25,0.45) {3};
\node[vblue] (c4) at (1.25,0.45) {4};

\draw[edgegray] (c5)--(c2);
\draw[edgered] (c1)--(c5);
\draw[edgered] (c3)--(c5);
\draw[edgered] (c4)--(c5);

%\node[anchor=west,font=\scriptsize,text=red!85!black] at (-1.65,-1.9)
%{red edges = sampled Schur-complement update};
\end{scope}

\end{tikzpicture}%
}
\caption{Dynamic dependency tracking. (a) Elimination tree, which pessimistically serializes $1 \rightarrow 3 \rightarrow 4 \rightarrow 5$. (b) Input graph. (c) After eliminating $0$, a sampled spanning tree (red) replaces the clique on its neighbors.}
\label{fig:mat_pattern}
\end{figure}
%--------------------------------------------------------------------
\subsection{Dependency Structure}

\Cref{fig:mat_pattern} illustrates why AC can expose more parallelism than complete Cholesky. Panel~(a) is the elimination tree of the graph in panel~(b), and a static schedule derived from it eliminates $1 \rightarrow 3 \rightarrow 4 \rightarrow 5$ in sequence. \texttt{ParAC} instead finds $1$, $3$, and $4$ ready at the same time once $0$ has been eliminated, as the trace later in this section shows. If vertex $0$ were eliminated by complete Cholesky, its neighbors would form a clique, inducing broad reachability and many dependencies. In contrast, AC forms only a sampled spanning tree, which reduces reachability and allows more vertices to proceed in parallel.

For \emph{exact} Cholesky on a statically known graph, the transitive dependencies encoded by the elimination tree are characterized by reachability.

\begin{lemma}[\cite{rosetarjan78,liu1990role}]
\label{lemma:reachability}
Define $j$ as a transitive dependency of $i$ if eliminating $j$ can affect $i$ through subsequent eliminations. Then $i$ depends on $j$ if and only if $j < i$ and there exists a path
\[
i \rightarrow p_1 \rightarrow p_2 \rightarrow \cdots \rightarrow p_n \rightarrow j
\]
such that every internal vertex satisfies $p_1,\ldots,p_n < i$.
\end{lemma}

\Cref{lemma:reachability} predicts elimination-tree ancestry from the \emph{input} graph alone. That is useful for exact factorization but inapplicable here, because under AC the active graph evolves stochastically, so the transitive paths it quantifies over are not known in advance. \texttt{ParAC} sidesteps the prediction entirely by reading readiness off the current active graph, where a single scalar suffices. Let
\[
\mathcal{G}^{(t)} \triangleq \bigl(V^{(t)},E^{(t)},m^{(t)}\bigr)
\]
be the active multigraph after $t$ elimination steps, where $m^{(t)}(i,j)$ denotes edge multiplicity. For an active vertex $i \in V^{(t)}$, define
\[
d_t(i) \triangleq \sum_{j<i} m^{(t)}(i,j),
\]
namely, the number of incident edges from lower-labeled active vertices, counted with multiplicity.

Readiness then follows directly from the invariant, with no appeal to reachability. Eliminating a pivot $k$ modifies exactly the vertices in $\mathcal{N}_k$, so an active vertex $i$ can still be modified only by the elimination of an active vertex $j<i$ that is \emph{currently} adjacent to $i$. If $d_t(i)=0$ there is no such $j$: every remaining coupling of $i$ is to a higher-labeled vertex, which will be eliminated after $i$, so $i$ may be eliminated immediately. If $d_t(i)>0$ then some active $j<i$ is adjacent to $i$, and eliminating $j$ will update $i$, so $i$ must wait. Hence an active vertex is ready exactly when $d_t(i)=0$.

The saving is that \texttt{ParAC} needs only the scalar counter $d_t(i)$, updated locally as edges are inserted and deleted, rather than any transitive reachability information. Because the criterion is evaluated on the active graph rather than predicted from the input, it remains exact under AC's stochastic fill-in, where a static schedule derived from \Cref{lemma:reachability} would be conservative.

%--------------------------------------------------------------------
\subsection{Proposed Tracking Strategy}

\texttt{ParAC} tracks dependencies online rather than estimating them statically. At any time, the ready set is
\[
Q_t \triangleq \{\, i \in V^{(t)} \mid d_t(i)=0 \,\},
\]
so vertices in $Q_t$ can be eliminated immediately and, subject to hardware availability, in parallel.

Initially, every input edge has multiplicity one, so
\[
d_0(i)=\left| \{\, j \mid e_{ij}\neq 0,\; j<i \,\} \right|.
\]
This is exactly the initialization of \texttt{dp[i]} in \Cref{a:cpu}. Vertices with $d_0(i)=0$ are inserted into the initial job queue, and the linked-list heads are initialized at the same time.

Each thread repeatedly polls its assigned queue locations and retrieves a ready pivot $k$ (\Cref{a:cpu}). This is the implementation-level realization of the ready set $Q_t$.

After gathering and sorting the active neighbors of $k$ (\Cref{a:cpu}), \texttt{ParAC} samples Schur-complement edges among those neighbors. Whenever a sampled edge is formed between active vertices $a$ and $b$, \texttt{ParAC} records the edge and increments the count of its larger-labeled endpoint,
$d(\max(a,b)) \leftarrow d(\max(a,b))+1$.
All sampled-edge insertions and increments complete before the pivot-edge decrements.

When the pivot $k$ is eliminated, all active edges incident to $k$ are removed. Thus, for every neighbor $i>k$,
$d(i) \leftarrow d(i)-m(i,k)$,
which appears in aggregate form as \texttt{dp[i]-=N\_k(i).multiplicity} in \Cref{a:cpu}.
After these insertion and deletion updates, any vertex whose dependency count becomes zero is inserted into the job queue. This is exactly the readiness condition
\[
d_t(i)=0 \iff i \in Q_t.
\]

\Cref{fig:mat_pattern} shows one example of the elimination process, following the order in which \Cref{a:cpu} performs the updates: insertions first, then deletions, with enqueueing triggered only by a positive-to-zero transition. Before eliminating vertex $0$, the input graph (panel~(b)) gives initial dependency counts $d=(d_0,d_1,d_2,d_3,d_4,d_5)=(0,1,0,1,1,2)$. First, the sampled Schur-complement edges $(1,5),(3,5),(4,5)$ are inserted (panel~(c)); each raises the count of its larger-labeled endpoint, so $d_5$ goes from $2$ to $5$. Only then is the pivot removed: deleting edges $(0,1),(0,3),(0,4),(0,5)$ decrements $d_1,d_3,d_4,d_5$ by one each, leaving $(d_1,d_2,d_3,d_4,d_5)=(0,0,0,0,4)$. Vertices $1$, $3$, and $4$ transition to zero and are enqueued; vertex $5$ does not.

This ordering is not incidental. A sampled edge raises the count of its larger-labeled endpoint, which is itself a neighbor of the pivot and therefore also due a decrement. Were the deletions applied first, such a vertex could transiently reach zero and be enqueued before its incoming sampled edge was recorded. Another thread could then begin eliminating a vertex that still had a pending dependency. Performing insertions first keeps the counter an upper bound on the outstanding dependencies at all times, so a zero crossing is always genuine.
 
Hence \texttt{ParAC} maintains the invariant that $d_t(i)$ equals the total multiplicity of lower-labeled incident edges of $i$ in the current active multigraph. A vertex becomes ready exactly when this counter reaches zero, at which point it is inserted into the ready queue.

This dependency rule is independent of the hardware backend. The CPU and GPU implementations in the next section differ in how updates are stored and searched, but both use the same counter-based readiness criterion and ready-queue semantics.

\paragraph{Distribution Preservation}
Distinct ready pivots are non-adjacent: if $i<j$ were both active and adjacent, then $j$ would have the lower-labeled active neighbor $i$, giving $d_t(j)>0$ and contradicting $j \in Q_t$.
Consequently, eliminating one ready pivot leaves every other ready pivot's incident star unchanged. Eliminating $j$ deletes the edges incident to $j$ and inserts sampled edges among $\mathcal{N}_j$; since $i \notin \mathcal{N}_j$, no edge incident to $i$ is deleted and no sampled edge can land on $i$. Because \texttt{SampleClique} draws using only the weights on the pivot's own star, each ready pivot therefore has the same conditional sampling distribution whether it is eliminated first, last, or concurrently.
The pivots' contributions need not be disjoint. Two ready pivots may share higher-labeled neighbors, and may even sample the same edge, which is why the active graph is a multigraph. But each contributes an additive rank-one term to the Schur complement, and addition commutes, so the accumulated result is independent of the order in which the contributions are applied.
With independent random draws, eliminating all of $Q_t$ simultaneously therefore produces the same joint distribution over the output factor $G$ as any sequential order respecting the dependency partial order.
In particular, for the same prescribed elimination order, the output distribution of \texttt{ParAC} matches that of a serial execution of the AC algorithm in Algorithms~\ref{a:rchol} and \ref{a:sample}~\cite{gao2023robust}, which is also implemented by \texttt{rchol} and is provably breakdown-free on connected graph Laplacians~\cite[Thm.~2.7, Cor.~2.8]{rchol}, \cite[Claim~3.5, Thm.~3.6]{gao2023robust}.
\section{Parallel Designs for CPUs and GPUs}
\label{sec:new_design}
We now present our parallelization method, \texttt{ParAC}, based on the dependency tracking strategy discussed above. We dissect how the architectural differences between CPUs and GPUs shape the design details of \texttt{ParAC} and present two algorithms, respectively. Both store the factor in the scaled form $L \approx F D F^{\top}$, where $F$ is unit lower triangular and $D$ is diagonal, which avoids a square root per column; this is the factor $G$ of \Cref{a:rchol} under $G = F D^{1/2}$.

At a high level, \texttt{ParAC} first selects an elimination ordering and labels all vertices accordingly. The rest of the elimination steps can be divided into three main stages, following \Cref{a:rchol} and \Cref{a:sample}:
\begin{enumerate}
  \item Search and organize fill-in updates for an eliminated vertex.
  \item Sort the neighbors of the pivot and perform sampling.
  \item Perform Schur-complement updates, update the dependencies, and schedule any vertex that is ready to be eliminated.
\end{enumerate}

Details are discussed in the subsequent subsections, with the flow chart previewed in \Cref{fig:ParAC_key}.

\begin{figure}[t]
  \centering
  \begin{adjustbox}{max width=\columnwidth}
  \begin{tikzpicture}[
    font=\Large,
    node distance=8mm and 12mm,
    box/.style={rounded corners, draw, thick, fill=black!3, inner sep=3.5pt, align=center},
    data/.style={draw, thick, fill=blue!0, inner sep=3.5pt, align=center},
    storeCPU/.style={draw, thick, fill=green!8, inner sep=3.5pt, rounded corners, align=center},
    storeGPU/.style={draw, thick, fill=blue!8, inner sep=3.5pt, rounded corners, align=center},
    lane/.style={draw, dashed, rounded corners, inner sep=6pt},
    annot/.style={inner sep=1pt},
    >=Latex
  ]

  %-------------------- Shared top row --------------------
  
  \node[data, minimum width=4.2cm] (counts) {\textbf{Dependency counts}\\ \(\texttt{dp}[v]\)};
  \node[data, right=of counts, minimum width=4.2cm] (ready) {\textbf{Ready set / queue}\\ \(\{\,v:\, \texttt{dp}[v]=0\,\}\)};
  \node[data, right=of ready, minimum width=4.2cm] (state) {\textbf{Active graph}\\ \(\mathcal{G}^{(t)} \rightarrow \mathcal{G}^{(t+1)}\)};

  \draw[->, thick] (counts) -- (ready);
  \draw[->, thick] (ready) -- (state);

  %-------------------- CPU lane (left) --------------------
  \node[box, below=20mm of counts, minimum width=7.2cm, minimum height=1.4cm, anchor=north] (cpuTitle)
    {\textbf{CPU}\\ one thread per vertex};
  \node[storeCPU, below=6mm of cpuTitle, minimum width=7.2cm] (cpuQueue)
    {\textbf{Task queue}\\ threads poll the ready queue};
  \node[box, below=6mm of cpuQueue, minimum width=7.2cm] (cpuElim)
    {\textbf{Reserve space}\\ allocate a chunk of the workspace \(O\)};
  \node[storeCPU, below=6mm of cpuElim, minimum width=7.2cm] (cpuBuf)
    {\textbf{Left-searching}\\ gather fill-ins from the linked list};
  \node[box, below=6mm of cpuBuf, minimum width=7.2cm] (cpuCommit)
    {\textbf{Eliminate \(v\)}\\ sort, sample Schur-complement edges, \\ store them in the reserved chunk};
  \node[box, below=6mm of cpuCommit, minimum width=7.2cm] (cpuDec)
    {\textbf{Update dependency} \(\texttt{dp}[\cdot]\)};

  % Downstream arrows (CPU lane)
  \draw[->, thick] (cpuQueue) -- (cpuElim);
  \draw[->, thick] (cpuElim) -- (cpuBuf);
  \draw[->, thick] (cpuBuf) -- (cpuCommit);
  \draw[->, thick] (cpuCommit) -- (cpuDec);

  %-------------------- GPU lane (right) --------------------
  \node[box, below=20mm of state, minimum width=7.2cm, minimum height=1.4cm, anchor=north] (gpuTitle)
    {\textbf{GPU}\\ one thread block per vertex};
  \node[storeGPU, below=6mm of gpuTitle, minimum width=7.2cm] (gpuKernel)
    {\textbf{Persistent kernel}\\ SMs poll the ready queue};
  \node[box, below=6mm of gpuKernel, minimum width=7.2cm] (gpuWarp)
    {\textbf{Reserve space}\\ allocate a chunk of the workspace \(O\)};
  \node[storeGPU, below=6mm of gpuWarp, minimum width=7.2cm] (gpuW)
    {\textbf{Right-searching}\\ look up the hashmap \(W\)};
  \node[box, below=6mm of gpuW, minimum width=7.2cm] (gpuSample)
    {\textbf{Eliminate \(v\)}\\ sort, sample Schur-complement edges, \\ store them at hashed locations};
  \node[box, below=6mm of gpuSample, minimum width=7.2cm] (gpuCommit)
    {\textbf{Update dependency} \(\texttt{dp}[\cdot]\)};

  % Downstream arrows (GPU lane)
  \draw[->, thick] (gpuKernel) -- (gpuWarp);
  \draw[->, thick] (gpuWarp) -- (gpuW);
  \draw[->, thick] (gpuW) -- (gpuSample);
  \draw[->, thick] (gpuSample) -- (gpuCommit);

  %-------------------- Fan-out from state (orthogonal, no overlap) --------------------
  \draw[->, thick]
    (state.south) -- ++(0,-8mm) -| ([xshift=6mm]cpuQueue.east) -- (cpuQueue.east);

  \draw[->, thick]
    (state.south) -- ++(0,-8mm) -| ([xshift=6mm]gpuKernel.east) -- (gpuKernel.east);

  %-------------------- Feedback edges (orthogonal, routed outside lanes) --------------------
  \draw[->, thick]
    (cpuDec.west) -- ++(-8mm,0) |- ([xshift=-8pt]counts.west);

\draw[->, thick]
  (gpuCommit.east) -- ++(18mm,0)
  |- ([yshift=12mm,xshift=-8pt]counts.north)
  -- ([xshift=-8pt]counts.north);

  %-------------------- Visual grouping boxes --------------------
  \node[lane, fit=(cpuTitle)(cpuQueue)(cpuElim)(cpuBuf)(cpuCommit)(cpuDec)] {};
  \node[lane, fit=(gpuTitle)(gpuKernel)(gpuWarp)(gpuW)(gpuSample)(gpuCommit)] {};

  % Labels
  \node[annot, above left=0mm and -2mm of cpuTitle.north west] {\textbf{CPU path}};
  \node[annot, above left=0mm and -2mm of gpuTitle.north west] {\textbf{GPU path}};

\end{tikzpicture}
  \end{adjustbox}
  \caption{Flow chart of the \texttt{ParAC} pipeline.}
  \label{fig:ParAC_key}
\end{figure}

%--------------------------------------------------------------------
\subsection{Common Design Details}
\label{subsec:challenges}
There are two considerations common to CPUs and GPUs. Our design starts from these common considerations.

\paragraph{Memory Layout}
A practical obstacle to an efficient parallelization is memory provisioning under randomization. Because AC randomly samples vertex pairs to form new edges, column-wise fill is unknown a priori. Rather than repeatedly resizing per-column containers, which is costly, we design data structures that avoid per-column reallocations and support concurrent producers.

Specifically, we pre-allocate a contiguous workspace $O$ that holds the final factor and temporary update data. Threads reserve column chunks by applying atomic \texttt{fetch\_add} to a shared global offset, which avoids repeated per-column reallocations as long as the workspace does not overflow. The CPU implementation allocates $O$ at five times the input nonzero count. On the GPU, $O$ and the additional open-addressed hashmap workspace $W$ use a configurable capacity that defaults to four times the input nonzero count and is increased to six times for \texttt{com-LiveJournal}.

\paragraph{Work Scheduling} 
%Since we dynamically check the readiness of each vertex during runtime, we need a mechanism to schedule the vertices once they are ready. 
A naive approach is to schedule the vertices once they are ready by using lock-related semantics, but this may cause severe thread contention that slows down the whole factorization.
We take the following approach instead: whenever a vertex becomes ready, it is inserted into a queue. All the threads (or thread blocks) poll their respective search indices inside the queue, which are assigned in a round-robin fashion based on the thread/block id. For example, if there are four threads, thread $0$ is assigned $\{0, 4, 8, \ldots\}$ while thread $1$ is assigned $\{1, 5, 9, \ldots\}$.

%--------------------------------------------------------------------
\subsection{CPU Algorithm}

\begin{algorithm}[!t]
\caption{\texttt{ParAC} on CPUs}
\label{a:cpu}
\begin{algorithmic}[1]
\Require Laplacian matrix $L \in \mathbb{R}^{N \times N}$, \texttt{num\_threads} threads
\Ensure Output array $O$ containing the triangular factor, diagonal matrix $D$

\State{$\texttt{dp}[i] \leftarrow |\{j \mid j < i,\, e_{ij} \neq 0\}|$ for all $i$}
\State{$q \leftarrow \{i \mid \texttt{dp}[i] = 0\}$;\quad $P \leftarrow$ linked-list heads}
\For{$\texttt{id} = \texttt{thread\_id}$ \textbf{to} $N-1$ \textbf{step} \texttt{num\_threads}}
    \State{$k \leftarrow q[\texttt{id}]$, spin-waiting if necessary}
    \State{Reserve a chunk of the workspace $O$ for $k$}
    \State{$\mathcal{N}_k \leftarrow$ original neighbor entries in column $k$ plus updates traversed from $P(k)$}
    \State{\textbf{if} $\mathcal{N}_k=\varnothing$ \textbf{then} $D(k,k)\leftarrow 0$; \textbf{continue}}
    \State{$D(k,k) \leftarrow \sum_{i \in \mathcal{N}_k} |\ell_{ik}|$}
    \State{Merge duplicates in $\mathcal{N}_k$, then sort by ascending $|\ell_{ik}|$}
    \State{$S \leftarrow$ suffix sums of $|\ell_{ik}|$ over $\mathcal{N}_k$}
    \For{$p = 1 : \left| \mathcal{N}_k \right| - 1$}
        \State{$u \leftarrow \operatorname{vertex}(\mathcal{N}_k[p])$}
        \State{Sample $v$ from $\mathcal{N}_k[p{+}1{:}]$ w.p. $|\ell_{kv}|/S[p{+}1]$}
        \State{$\operatorname{AtomicAdd}(\texttt{dp}[\max(u,v)],1)$}
        \State{$P(\min(u,v)) \leftarrow -\frac{S[p+1]\,\ell_{ku}}{\ell_{kk}}\,\bb_{uv}\bb_{uv}^{\top}$}
    \EndFor
    \State{Store the normalized factor column for $k$ in $O$}
    \For{$p = 1 : |\mathcal{N}_k|$}
        \State{$u, m \leftarrow \operatorname{vertex}(\mathcal{N}_k[p]), \operatorname{multiplicity}(\mathcal{N}_k[p])$}
        \State{$r \leftarrow \operatorname{AtomicSub}(\texttt{dp}[u],m)$}
            \Comment{prior count}
        \State{\textbf{if} $r=m$ \textbf{then} $\operatorname{AtomicAppend}(q,u)$}
    \EndFor
\EndFor
\end{algorithmic}
\end{algorithm}

The CPU implementation is based on the ``left-searching'' mechanism: at each step, the Schur-complement updates are stored locally and chained together by a linked list. Vertices to be eliminated later will traverse the list (hence the term ``left'') and aggregate these updates.
This is analogous to the left-looking strategy in sparse LU factorization~\cite{superlu-overview}, where a column aggregates contributions from all previously factored columns before it is processed.

\paragraph{Reserve Space and Left-Searching}
Consider a batch $B \triangleq \{v_1,\ldots,v_m\}$ of pivots processed concurrently after they are obtained from the job queue. For each pivot $v_k$, we estimate the space required for its completed column from its original neighbor entries and pending-update count, then atomically advance the shared offset; the old value gives the chunk's starting index. We gather the pending updates by left-searching through the linked list, merge them with the original entries, and write the completed column to the reserved chunk. Since each neighbor generates at most one sampled edge, the storage required for a pivot remains linear in its gathered neighborhood size.

\paragraph{Eliminate and Schedule}
We then aggregate the fill-ins. Let $T \triangleq \{t_1, \ldots, t_q\}$ be the set modified by the Schur-complement update of $B$ and let $P \triangleq \{p_1 \rightarrow \texttt{fill-ins}, \ldots, \\ p_q \rightarrow \texttt{fill-ins}\}$ represent the pointers owned by $T$, where \texttt{fill-ins} refer to the existing fill-ins each vertex $t_h \in T$ must aggregate. We use fill-ins to indicate all entries, even if such an entry already exists (in which case, we merge them). Suppose $v_k \in B$ modifies $t_h$ via the Schur-complement update. Then, $v_k$ inserts the sample it generates into $t_h$'s linked list. That is, $p_h \rightarrow \texttt{sample}(v_k) \rightarrow \texttt{fill-ins}$, where $\texttt{sample}(v_k)$ is generated by some neighbor of $v_k$ and is stored in the local chunk owned by $v_k$.

Since $B$ is eliminated in parallel, a race condition may happen if multiple elements in $B$ update the same element in $T$. %A simple and efficient solution to this problem is to 
We use atomic exchanges to preserve the integrity of the linked list. Sampled edges and their dependency increments are recorded before pivot-edge decrements; vertices that then reach zero are enqueued, exactly as specified in \Cref{parallel:insight}.

The overall procedure is summarized in \Cref{a:cpu}.

%--------------------------------------------------------------------
\subsection{GPU Algorithm}

\begin{algorithm}[!t]
\caption{\texttt{ParAC} on GPUs}
\label{a:gpu}
\begin{algorithmic}[1]
\Require Laplacian matrix $L \in \mathbb{R}^{N \times N}$, \texttt{num\_blocks} thread blocks
\Ensure Output array $O$ containing the triangular factor, diagonal matrix $D$

\State{$\texttt{dp}[i] \leftarrow |\{j \mid j < i,\, e_{ij} \neq 0\}|$ for all $i$}
\State{$q \leftarrow \{i \mid \texttt{dp}[i] = 0\}$;\quad $W \leftarrow$ empty update workspace}
\For{$\texttt{id} = \texttt{block\_id}$ \textbf{to} $N-1$ \textbf{step} \texttt{num\_blocks}}
    \State{$k \leftarrow q[\texttt{id}]$, spin-waiting if necessary}
    \State{Reserve a chunk of the workspace $O$ for $k$}
    \State{$\mathcal{N}_k \leftarrow$ original neighbor entries in column $k$ plus updates found by probing $W$ from $\texttt{hash}(k)$}
    \State{\textbf{if} $\mathcal{N}_k=\varnothing$ \textbf{then} $D(k,k)\leftarrow 0$; \textbf{continue}}
    \State{$D(k,k) \leftarrow \sum_{i \in \mathcal{N}_k} |\ell_{ik}|$}
    \State{Merge duplicates in $\mathcal{N}_k$ by sort and prefix sum, then sort by ascending $|\ell_{ik}|$}
    \State{$S \leftarrow$ parallel suffix sums of $|\ell_{ik}|$ over $\mathcal{N}_k$}
    \For{$p = 1 : \left| \mathcal{N}_k \right| - 1$ \textbf{in parallel}}
        \State{$u \leftarrow \operatorname{vertex}(\mathcal{N}_k[p])$}
        \State{Sample $v$ from $\mathcal{N}_k[p{+}1{:}]$ w.p. $|\ell_{kv}|/S[p{+}1]$}
        \State{$\operatorname{AtomicAdd}(\texttt{dp}[\max(u,v)],1)$}
        \State{$W(\texttt{hash}(\min(u,v))) \leftarrow -\frac{S[p+1]\,\ell_{ku}}{\ell_{kk}}\,\bb_{uv}\bb_{uv}^{\top}$}
    \EndFor
    \State{Store the normalized factor column for $k$ in $O$}
    \For{$p = 1 : |\mathcal{N}_k|$ \textbf{in parallel}}
        \State{$u, m \leftarrow \operatorname{vertex}(\mathcal{N}_k[p]), \operatorname{multiplicity}(\mathcal{N}_k[p])$}
        \State{$r \leftarrow \operatorname{AtomicSub}(\texttt{dp}[u],m)$}
            \Comment{prior count}
        \State{\textbf{if} $r=m$ \textbf{then} $\operatorname{AtomicAppend}(q,u)$}
    \EndFor
\EndFor
\end{algorithmic}
\end{algorithm}

\label{sec:gpu_alg}
GPU design is considerably more involved. First, \texttt{ParAC} must run in units of warps to avoid warp divergence, whereas on the CPU the smallest unit is a single thread. Second, AC's sparse sampling yields low computational intensity per elimination step, so a naively launched kernel may not amortize its launch overhead. Third, inter-block synchronization is limited and error-prone, and kernels coordinating across blocks can deadlock. Fourth, preserving the weight-sorted AC sampling rule in \Cref{a:sample} requires sorting each dynamically sized neighborhood.

All these considerations prompt us to use a different mechanism to implement \texttt{ParAC} on GPUs. We use ``right-searching,'' in which Schur-complement updates are written near a target-specific hash location. A later pivot can then gather them by parallel probing rather than by traversing a serial linked list.

\paragraph{Address Warp Divergence}
We assign each job to a thread block consisting of warp(s). Then, we structure the computation in a manner that takes advantage of multi-threading within each warp, by mapping the neighbors of each vertex to a thread. There may be idle threads, if the number of neighbors is fewer than $32$, or we may need multiple rounds of coverage if there are more neighbors than threads.

\paragraph{Reduce Kernel Launches}
In contrast to the subtree-streaming exact Cholesky approach of Rennich et al.~\cite{gpu_sparse_chol}, we adopt a persistent kernel. All blocks remain active and continuously poll the queue at their assigned locations, fetching and performing a job whenever one is available. We repeat this in a loop until the factorization is complete. This strategy fits the entire factorization, minus some minor preprocessing, into one kernel, thereby reducing kernel launch costs.

\paragraph{Reserve Space and Right-Searching}
Linked list-based design, as in the CPU case, is no longer practical, because pointer jumping cannot take advantage of intra-block parallelism. To let a vertex $v_k \in B$ efficiently find its fill-ins, we cluster them near a target-specific starting location. This motivates a linear-probing, array-based hashmap design in which elements are inserted in groups. We call this hashmap array $W$ and refer to future searches conducted in this map as ``right-searching,'' because Schur-complement updates are written near the target column's hash location rather than stored in a linked list.

More specifically, the vertex $v_k$ generates a hashcode $\texttt{hash}(v_k)$, which indicates the initial search location. The block of threads responsible for eliminating $v_k$ will then search the array in parallel until it finds the expected number of fill-ins. The workspace $W$ differs from $O$: $W$ only temporarily stores fill-ins, and the entries holding $v_k$'s fill-ins are freed for reuse once they are moved to $O$. Each entry of $W$ is free, busy, or occupied. Search threads retry a busy entry, whereas insertion threads continue probing until they claim a free entry.

\paragraph{Sorting}
We merge fill-ins by sorting and prefix-sum. NVIDIA's CUB provides efficient warp-level sorting, but requires the elements per thread to be known at compile time, which the dynamically varying fill-in count violates. We therefore wrote block-level odd-even and bitonic sorts handling an arbitrary number of elements, choosing between CUB and ours by an element-count threshold. Our sorter pads the array with sentinel values to the next power of two, matching the network's fixed-radix structure.

\paragraph{Eliminate and Schedule}
Consider the set of vertices updated by $v_k$'s Schur-complement update (i.e., $\mathcal{U} \triangleq \{t_h \mid t_h \in T,\ t_h \in \mathcal{N}(v_k)\}$). The thread block calculates a value \texttt{hash}$(a) + \texttt{fill\_in\_count}(a)$ for every $a \in \mathcal{U}$ and inserts them into the corresponding location in parallel, where $\texttt{fill\_in\_count}(a)$ refers to the number of existing fill-ins of $a$. Adding this value can potentially speed up insertion in most cases, since the spots before \texttt{hash}$(a) + \texttt{fill\_in\_count}(a)$ are likely taken. The dependency calculation and queue scheduling are similar to those of the CPU algorithm.

\paragraph{Efficient Hashing}
Hashing quality has a significant impact on GPU performance because nearby starting locations can cause linear-probing conflicts. Let $B$ be the pivots active concurrently at some instant, and let $T_B$ be the distinct target vertices receiving their updates. For the base-slot map $h_\sigma:V\rightarrow\{0,\ldots,|W|-1\}$ induced by a vertex permutation $\sigma$, and for $|T_B| \geq 2$, a useful placement proxy is the smallest circular gap
\begin{equation*}
    \begin{aligned}
        \delta_\sigma(T_B)
        &\triangleq \min_{\substack{a,b \in T_B \\ a \neq b}}
        d_W\bigl(h_\sigma(a),h_\sigma(b)\bigr), \\
        d_W(x,y)
        &\triangleq \min\bigl(|x-y|,\; |W|-|x-y|\bigr).
    \end{aligned}
\end{equation*}
Here $d_W$ accounts for wraparound in the workspace. A larger $\delta_\sigma(T_B)$ tends to reduce overlap between probing sequences, but it is not a complete model of probing cost, which also depends on per-target update counts and the current occupancy of $W$. Because target sets vary during factorization and with sampled fill, no fixed mapping maximizes this score for every set. We therefore use a fixed random permutation of the vertex labels to assign base slots; this heuristic performs better empirically than the identity mapping and is separate from the elimination order.

With all the addressed details, the overall GPU implementation is summarized in \Cref{a:gpu}.

%--------------------------------------------------------------------
\section{Numerical Results}
\label{sec:experiments}

\paragraph{Experimental Methodology}
For each benchmark, we solve a linear system $Lx = b$. The right-hand sides are generated directly from uniform random samples and mean-centered, ensuring $b\perp\mathbf{1}$ for singular graph Laplacians. \texttt{ParAC} computes an IC factorization $L \approx GG^\top$, where $G$ is sparse and lower triangular. We then use this factor as a preconditioner for the preconditioned conjugate gradient (PCG) method, which returns an approximate solution $\tilde{x}$. We report three metrics:

\begin{enumerate}
\item \textbf{Factorization Time}: The wall-clock time required to construct the approximate Cholesky factor $G$.
\item \textbf{Solve Time}: The wall-clock time for the PCG solver to converge to the target tolerance.
\item \textbf{Iteration Count}: The number of PCG iterations required to reach a relative residual on the order of $10^{-6}$.
%
%\item \textbf{Relative Residual}: The final error calculated as
%\[
%\|L\tilde{x} - b\|_2 / \|b\|_2.
%\]
%
%\item \textbf{Storage Ratio}: 
\end{enumerate}

Because \texttt{ParAC} is randomized, individual timings and iteration counts may vary slightly between runs; the tables report the measured runs used in our comparison.
For the four \texttt{ichol}-preconditioned CPU solves that exceeded the initial 1,000-iteration cap, we raised the cap to 10,000 and report the converged measurements.
We store matrices using the standard Laplacian/SDDM sign convention, with positive diagonal and nonpositive off-diagonal entries. For graph inputs supplied as nonnegative adjacency weights, preprocessing negates the edge weights and constructs the diagonal when it is absent.
Some matrices in our suite are SDDM matrices rather than pure graph Laplacians. For SDDM inputs with nonzero row sums, preprocessing appends a ground node to form an augmented graph Laplacian~\cite{gremban1996combinatorial}. After factoring this augmented system, we remove the ground-node row and column from the factor and use the remaining factor to precondition the original SDDM system.

\paragraph{Ordering for \texttt{ParAC}}
As in exact Cholesky factorization, the performance of \texttt{ParAC} depends on the initial ordering, that is, on the permutation applied to the input matrix. Minimizing fill-in is the primary objective in the sequential setting, but the ordering also determines how much parallelism is available. We evaluate three representative strategies:
\begin{enumerate}
\item \textbf{Approximate Minimum Degree (AMD)}~\cite{amd-order}: A widely used fill-reducing heuristic, and an effective one for \texttt{rchol}~\cite{rchol}.
\item \textbf{NNZ-Sorted}: Columns are ordered by ascending degree, where the degree of a column is its number of non-zero (NNZ) entries. Ties are broken randomly.
\item \textbf{Random}: A uniformly random ordering, adopted in~\cite{kyng2016approximate} to make the theoretical analysis tractable.
\end{enumerate}
Orderings designed specifically for randomized elimination have also been proposed~\cite{powerrchol}. We restrict our study to the three general-purpose heuristics above, none of which requires domain knowledge of the input.

\paragraph{Test Matrices}
\Cref{table:inputs} lists the number of columns and the number of non-zeros (NNZ) for each matrix in our test suite, which falls into four groups:
\begin{itemize}
\item \textbf{General Scientific Computing}: Five matrices from the SuiteSparse Matrix Collection~\cite{suitesparse} representing CFD, ecology, structural mechanics, and circuit simulation.
\item \textbf{Structured 3D Poisson}: Three matrices generated with the \texttt{Laplacians.jl} package~\cite{laplaciansjl}. Each is a finite-difference discretization of the Poisson equation on a regular 3D grid with the standard 7-point stencil. The three differ in their coefficients: uniform, anisotropic, and high-contrast, abbreviated \texttt{uniform}, \texttt{aniso}, and \texttt{contrast Poisson} in the results tables.
\item \textbf{Porous Media Flow}: The \texttt{spe16m} matrix, derived from the Society of Petroleum Engineers SPE10 benchmark~\cite{spe}, is a large-scale industrial model of fluid flow in porous media.
\item \textbf{Graph Laplacians}: Six network matrices from SuiteSparse~\cite{suitesparse}. These are adjacency matrices of weighted undirected graphs. We replace each off-diagonal weight by its negative absolute value and set each diagonal entry to the sum of the absolute off-diagonal weights in its row.
\end{itemize}

% \begin{table}[h!]
% \centering
% \caption{Properties of the test matrices. 
% }
% \label{table:inputs}
% \begin{tabular}{lrr}
% \toprule
% Matrix Name & \# Columns & \# Nonzeros \\
% \midrule
% parabolic\_fem & 525,825 & 3,674,625 \\
% ecology1 & 1,000,000 & 4,996,000 \\
% ecology2 & 999,999 & 4,995,991 \\
% apache2 & 715,176 & 4,817,870 \\
% G3\_circuit & 1,585,478 & 7,660,826 \\
% \midrule
% uniform 3D poisson & 14,348,907 & 100,088,055 \\
% anisotropic 3D poisson & 14,348,907 & 100,088,055 \\
% high-contrast 3D poisson & 14,348,907 & 100,088,055 \\
% \midrule
% spe16m & 16,003,008 & 111,640,032 \\
% \midrule
% GAP-road & 23,947,347 & 57,708,624 \\
% com-LiveJournal & 3,997,962 & 69,362,378 \\
% europe\_osm & 50,912,018 & 108,109,320 \\
% delaunay\_n24 & 16,777,216 & 100,663,202 \\
% venturiLevel3 & 4,026,819 & 16,108,474 \\
% belgium\_osm & 1,441,295 & 3,099,940 \\
% \bottomrule
% \end{tabular}
% \end{table}

\begin{table}[h!]
\centering
\caption{Properties of the test matrices.}
\label{table:inputs}
\begin{tabular}{lrrr}
\toprule
Matrix Name & \# Columns & NNZ & NNZ / Col \\
\midrule
parabolic\_fem & 525,825 & 3,674,625 & 6.99 \\
ecology1 & 1,000,000 & 4,996,000 & 5.00 \\
ecology2 & 999,999 & 4,995,991 & 5.00 \\
apache2 & 715,176 & 4,817,870 & 6.74 \\
G3\_circuit & 1,585,478 & 7,660,826 & 4.83 \\
\midrule
uniform 3D Poisson & 14,348,907 & 100,088,055 & 6.98 \\
anisotropic 3D Poisson & 14,348,907 & 100,088,055 & 6.98 \\
high-contrast 3D Poisson & 14,348,907 & 100,088,055 & 6.98 \\
\midrule
spe16m & 16,003,008 & 111,640,032 & 6.98 \\
\midrule
GAP-road & 23,947,347 & 57,708,624 & 2.41 \\
com-LiveJournal & 3,997,962 & 69,362,378 & 17.35 \\
europe\_osm & 50,912,018 & 108,109,320 & 2.12 \\
delaunay\_n24 & 16,777,216 & 100,663,202 & 6.00 \\
venturiLevel3 & 4,026,819 & 16,108,474 & 4.00 \\
belgium\_osm & 1,441,295 & 3,099,940 & 2.15 \\
\bottomrule
\end{tabular}
\end{table}

\begin{figure}[h!]
  \centering
  \includegraphics[height=1.0\linewidth,width=0.75\linewidth]{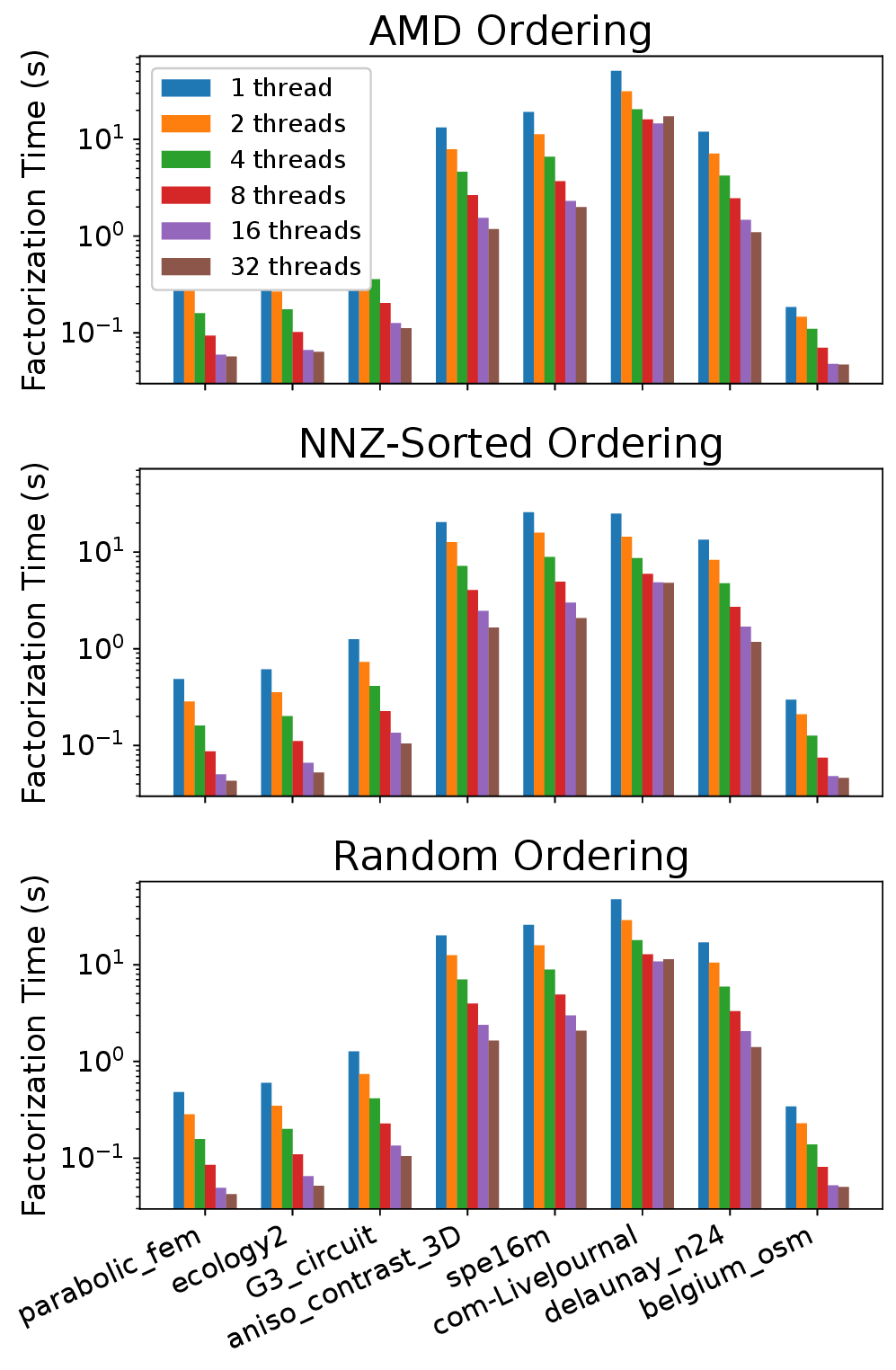}
  
  \caption{CPU factorization time under three orderings.}
  \label{experiment:cpu-scaling}
\end{figure}
% \begin{table}[t]
% \label{experiment:matrix-set}
% \centering
% \caption{Test matrices.}
% \label{table:inputs}
% \begin{tabular}{lrrc}
% \toprule
% Matrix Name & \# Columns & \# Nonzeros & \\
% \midrule
% parabolic\_fem & 525,825 & 3,674,625 & \multirow{5}{*}{engineering} \\
% ecology1 & 1,000,000 & 4,996,000 & \\
% ecology2 & 999,999 & 4,995,991 &  \\
% apache2 & 715,176 & 4,817,870 &  \\
% G3\_circuit & 1,585,478 & 7,660,826 &  \\
% \hline
% uniform 3D poisson & 14,348,907	& 100,088,055 & \\
% anisotropic 3D poisson & 14,348,907	& 100,088,055 &  \\
% high contrast 3D poisson & 14,348,907	& 100,088,055 &  \\
% \hline
% spe16m & 16,003,008 &	111,640,032 &  \\
% \hline
% GAP-road & 23,947,347 & 57,708,624 & \multirow{6}{*}{networks} \\
% com-LiveJournal & 3,997,962 & 69,362,378 &  \\
% delaunay\_n24 & 16,777,216 & 100,663,202 &  \\
% venturiLevel3 & 4,026,819 & 16,108,474 &  \\
% europe\_osm & 50,912,018 & 108,109,320 &   \\
% belgium\_osm & 1,441,295 & 3,099,940 &   \\
% \bottomrule
% \end{tabular}
% \end{table}

\paragraph{Reference Baselines}
We compare \texttt{ParAC} against deterministic incomplete factorizations and AMG solvers. For IC we use MATLAB on the CPU and NVIDIA \texttt{cuSPARSE}~\cite{cusparse} on the GPU. For AMG we use BoomerAMG~\cite{HensonYang2002} through the \texttt{hypre} library~\cite{falgout2002hypre} on the CPU and NVIDIA \texttt{AmgX}~\cite{naumov2015amgx} on the GPU.
For the 15-matrix comparison, we use two standard \texttt{AmgX} solver configurations; the complete settings and per-matrix assignment are provided in the accompanying code repository.
Applying the preconditioner $GG^\top$ requires triangular solves in the iterative phase, for both \texttt{ParAC} and IC. We perform these with Intel MKL on the CPU and \texttt{cuSPARSE} on the GPU, respectively.

\paragraph{Hardware Platform}
We ran all experiments on the NERSC Perlmutter supercomputer, on compute nodes with \textbf{AMD EPYC 7763 CPUs} and \textbf{NVIDIA A100 GPUs with 40\,GB of HBM2 memory}~\cite{nersc_perlmutter}. Unless otherwise noted, CPU \texttt{ParAC} factorization and \texttt{hypre} use 32 threads; the \texttt{ParAC} and \texttt{ichol} triangular solves use the same MKL thread configuration. On the GPU, \texttt{ParAC} uses a default configuration of 512 blocks with one warp per block. The exception is \texttt{com-LiveJournal}, for which we use 96 blocks with 8 warps
per block. Its high average degree makes the per-vertex workload much larger, and the wider configuration is needed to absorb it.

%--------------------------------------------------------------------
\subsection{CPU Results}

\paragraph{Impact of Ordering on CPU Performance}
We evaluate the impact of the three ordering heuristics, AMD, NNZ-sorted, and random, on the factorization phase. \Cref{experiment:cpu-scaling} shows a representative subset of eight benchmarks; the summary below uses the full 15-matrix suite.

Scaling from 1 to 32 threads gives roughly $5$--$11\times$ speedup on most matrices, with smaller gains on \texttt{com-LiveJournal} and \texttt{belgium\_osm}. AMD generally has the lowest low-thread factorization cost, although NNZ-sorted or random can overtake it at 32 threads. \texttt{com-LiveJournal} is the largest reversal: AMD takes $17.3$\,s versus $4.8$\,s for NNZ-sorted and gains only $2.9\times$ from 1 to 32 threads. This dense, irregular graph illustrates the trade-off between fill reduction and parallelism~\cite{liu1989reorder}.

Prior \texttt{rchol} results found more localized nonzero patterns under AMD, which can improve cache behavior during PCG~\cite{rchol}; together with our timings, this supports AMD as the default CPU ordering. The full results nevertheless show that NNZ-sorted can be preferable on dense inputs or at high thread counts.

\begin{table*}[t] 
    \centering
    \caption{CPU performance: \texttt{ParAC} vs. MATLAB \texttt{ichol} (threshold-based) and \texttt{hypre} (BoomerAMG).
    \colorbox{pink!50}{Pink}: iterations (randomized vs. deterministic IC).
    \colorbox{cyan!15}{Cyan}: factorization/setup cost.
    \colorbox{lightgray}{Gray}/\colorbox{lightgray!40}{light gray}: \texttt{ParAC} beats/matches \texttt{hypre}.
    } 
    \label{experiment:cpu} 

    \footnotesize
    \begin{tabular}{l 
                    >{\columncolor{cyan!15}}
                    S[table-format=2.2]     
                    S[table-format=3.2]     
                    S[table-format=3.2] % New Total Column ParAC
                    >{\columncolor{pink!50}}
                    S[table-format=3.0, scientific-notation=false, round-mode=none] 
                    S[table-format=3.2]     
                    S[table-format=4.2]     
                    S[table-format=4.2] % New Total Column ichol
                    >{\columncolor{pink!50}}
                    S[table-format=4.0, scientific-notation=false, round-mode=none] 
                    >{\columncolor{cyan!15}}
                    S[table-format=3.2, round-mode=places, round-precision=2]     
                    S[table-format=2.2, round-mode=places, round-precision=2]     
                    S[table-format=3.2] % New Total Column hypre
                    S[table-format=2.0, scientific-notation=false, round-mode=none] 
                   }
        \toprule
        Matrix Name & \multicolumn{4}{c}{\texttt{ParAC}} & \multicolumn{4}{c}{MATLAB \texttt{ichol}} & \multicolumn{4}{c}{\texttt{hypre}} \\
        \cmidrule(lr){2-5} \cmidrule(lr){6-9} \cmidrule(lr){10-13} 
         & {\shortstack{Fact. \\ Time (s)}} 
         & {\shortstack{Solve \\ Time (s)}}
         & {\shortstack{Total \\ Time (s)}}    
         & {\shortstack{Iter. \\ No.}}          
         & {\shortstack{Fact. \\ Time (s)}} 
         & {\shortstack{Solve \\ Time (s)}}
         & {\shortstack{Total \\ Time (s)}}    
         & {\shortstack{Iter. \\ No.}}
         & {\shortstack{Setup \\ Time (s)}} 
         & {\shortstack{Solve \\ Time (s)}}
         & {\shortstack{Total \\ Time (s)}}    
         & {\shortstack{Iter. \\ No.}} \\     
        \midrule
        parabolic\_fem    & 0.06 & 0.66  & 0.72  & 36   & 0.23 & 3.46    & 3.69    & 231  & 0.25 & 0.16  & 0.41  & 7    \\
        ecology1          & 0.06 & 1.09  & 1.15  & 42   & 0.13 & 16.12   & 16.25   & 637  & 0.39 & 0.23  & 0.62  & 7    \\
        ecology2          & 0.06 & 1.11  & 1.17  & 43   & 0.13 & 16.19   & 16.32   & 844  & 0.41 & 0.23  & 0.64  & 7    \\
        apache2           & 0.12 & 0.73  & 0.85  & 31   & 0.40 & 4.71    & 5.11    & 225  & 0.32 & 0.29  & 0.61  & 9    \\
        G3\_circuit       & 0.11 & 2.19  & 2.30  & 48   & 0.43 & 9.43    & 9.86    & 222  & 0.68 & 0.48  & 1.16  & 8    \\
        %\rowcolor{lightgray!40}
        uniform Poisson   & 2.61 & 16.74 & 19.35 & 30   & 8.96 & 61.62   & 70.58   & 102  & 9.37 & 5.47  & 14.84 & 8    \\ 
        \rowcolor{lightgray}
        aniso Poisson     & 1.18 & 5.59  & 6.77  & 11   & 5.17 & 4.53    & 9.70    & 7    & 4.22 & 4.66  & 8.88  & 6    \\ 
        contrast Poisson  & 1.57 & 75.67 & 77.24 & 142  & 6.70 & 56.35   & 63.05   & 91   & 8.23 & 5.50  & 13.73 & 8    \\ 
        spe16m            & 2.00 & 30.14 & 32.14 & 53   & 7.76 & 55.74   & 63.50   & 85   & 8.52 & 7.16  & 15.68 & 9    \\ 
        GAP-road          & 0.83 & 39.65 & 40.48 & 71   & 1.68 & 1727.31 & 1728.99 & 2285 & 13.28& 13.39 & 26.67 & 13   \\ 
        \rowcolor{lightgray}
        com-LiveJournal   & 17.32& 17.83 & 35.15 & 23   & 193.48& 14.14  & 207.62  & 15   & 236.16& 17.46& 253.62& 18   \\ 
        europe\_osm       & 1.67 & 85.22 & 86.89 & 72   & 2.70 & 2752.78 & 2755.48 & 2008 & 31.71& 33.85& 65.56 & 15   \\ 
        \rowcolor{lightgray!40}
        delaunay\_n24     & 1.10 & 18.69 & 19.79 & 33   & 5.81 & 1128.34 & 1134.15 & 1701 & 10.09& 7.43 & 17.52 & 10   \\ 
        venturiLevel3     & 0.21 & 5.50  & 5.71  & 51   & 0.90 & 168.88  & 169.78  & 1458 & 1.95 & 1.64  & 3.59  & 9    \\ 
        \rowcolor{lightgray!40}
        belgium\_osm      & 0.05 & 1.37  & 1.42  & 43   & 0.08 & 7.08    & 7.16    & 215  & 0.63 & 0.60  & 1.23  & 11   \\ 
        \bottomrule
    \end{tabular}
\end{table*}

\paragraph{Comparison with Baselines}
We compare the wall-clock performance of \texttt{ParAC} against two baselines: threshold-based incomplete Cholesky (\texttt{ichol}) from MATLAB, and the AMG solver in the \texttt{hypre} library. \Cref{experiment:cpu} reports the results.

\textbf{Performance vs. \texttt{ichol}}:
Both \texttt{ParAC} and \texttt{ichol} are incomplete factorization methods, so the two are direct competitors. For an iso-memory comparison, we tuned the drop tolerance on each benchmark so that the \texttt{ichol} factor occupies as much memory as the \texttt{ParAC} factor, or slightly more. \texttt{ParAC} is faster than \texttt{ichol} on 14 of the 15 test cases. The one exception is \texttt{contrast Poisson}, where \texttt{ParAC} is about 20\% slower. 

\texttt{ParAC} uses fewer iterations than \texttt{ichol} on 12 of the 15 benchmarks; the exceptions are \texttt{aniso Poisson}, \texttt{contrast Poisson}, and \texttt{com-LiveJournal}. On several matrices, including \texttt{ecology1}\footnote{\texttt{ecology1} is an ill-conditioned, nearly singular landscape-ecology matrix.} and \texttt{GAP-road}, the total execution time of \texttt{ParAC} is more than $10\times$ lower than that of \texttt{ichol}; on \texttt{com-LiveJournal} the margin is $5.9\times$. The large gaps on \texttt{ecology1}, \texttt{GAP-road}, and similar cases come from the iteration counts of \texttt{ichol}, not from MATLAB overhead: factorization accounts for at most $1\%$ of \texttt{ichol}'s total time, the PCG solve accounts for the rest, and the two methods cost within $1.4\times$ of each other per iteration. On the network problems where \texttt{ParAC} does use fewer iterations, the count for \texttt{ichol} is up to $50\times$ higher. \texttt{com-LiveJournal} is different: its speedup comes from much cheaper factorization despite the higher \texttt{ParAC} iteration count (\Cref{experiment:cpu}).

\textbf{Performance vs. BoomerAMG (\texttt{hypre})}\footnote{We used the default parameter configuration.}:
AMG is highly optimized for many mesh-based PDE discretizations, where a well-matched multilevel hierarchy can yield convergence largely independent of mesh refinement~\cite{RugeStuben1987,HensonYang2002}. The setup and factorization columns of \Cref{experiment:cpu} report the principal preconditioner-construction cost for each method. In the reported measurements, \texttt{hypre}'s setup exceeds \texttt{ParAC}'s factorization on all 15 benchmarks, and by more than an order of magnitude on several network problems. In total time, however, \texttt{hypre} is generally the faster solver. It wins on uniform and high-contrast Poisson, and also on \texttt{GAP-road}, \texttt{europe\_osm}, and \texttt{belgium\_osm}, where its low iteration counts offset the higher construction cost. \texttt{ParAC} is substantially faster on \texttt{com-LiveJournal} ($7.2\times$), the densest matrix in the suite, and modestly faster on \texttt{aniso Poisson} (both highlighted in gray).

The advantage of \texttt{ParAC} over AMG therefore lies in robustness and construction cost, not in total time across the board. Structured discretizations generally favor AMG, although anisotropic Poisson is an exception in our suite. \Cref{sec:chimera} tests the robustness side of the trade directly on heterogeneous SDDM systems.

The reported factorization-cost margin also leaves headroom that we do not spend here. An AC variant that invests more in construction could cut the iteration count and close the total-time gap. AC($k$)~\cite{gao2023robust} is the natural candidate. AC($2$) uses two samples per original entry, trading a denser, more costly factor for the possibility of fewer PCG iterations. Whether that reduction repays the higher construction and application cost is untested, and we leave it to future work.

%Moreover, \texttt{ParAC} is competitive with \texttt{hypre} on problems from the scientific domains, while being superior than \texttt{hypre} on graph problems (rows 6--11). AMG preconditioners like \texttt{hypre} have a natural advantage for mesh-based problems; for graphs with irregular sparsity patterns and varied densities, they are less competitive. \texttt{ParAC} shines in the latter case.

%It is worth noting that using the randomized preconditioner \texttt{ParAC}, the iterative solver generally converges faster than using the non-randomized \texttt{ichol}. In our tests, the right-hand side $b$ is generated by using a Gaussian random $x$. Such a $b$ has a higher variance for the components along the top eigenvector directions. Randomized preconditioners are robust to this situation.
\begin{table*}
  \centering
  \caption{GPU performance: \texttt{ParAC} vs. \texttt{cuSPARSE} IC(0) (zero fill-in) and \texttt{AmgX}.
  We report only a total for IC(0), since its factorization time is a negligible fraction of that total.
  \texttt{ParAC} totals include the mandatory \texttt{cuSPARSE} symbolic analysis.
  \textcolor{red}{Red}: reached the $10^4$-iteration cap or OOM.
  \colorbox{pink!50}{Pink}: iterations; \colorbox{cyan!15}{cyan}: factorization/setup;
  \colorbox{lightgray}{gray}/\colorbox{lightgray!40}{light gray}: \texttt{ParAC} beats/matches \texttt{AmgX}.
  }
    \label{experiment:gpu}
    \footnotesize
  \begin{tabular}{l
      >{\columncolor{cyan!15}}
      S[table-format=5.0]
      S[table-format=4.0]
      S[table-format=5.0]
      >{\columncolor{pink!50}}
      S[table-format=3.0]
      %@{\hspace{0.3cm}}
      S[table-format=6.0]
      >{\columncolor{pink!50}}
      S[table-format=5.0]
      %@{\hspace{0.3cm}}
      >{\columncolor{cyan!15}}
      S[table-format=4.0] 
      S[table-format=5.0]
      S[table-format=5.0]
      S[table-format=2.0]
      }
    \toprule
    Matrix Name & \multicolumn{4}{c}{\texttt{ParAC}} & \multicolumn{2}{c}{\texttt{cuSPARSE} IC(0)} & \multicolumn{4}{c}{\texttt{AmgX}} \\
    \cmidrule(lr){2-5} \cmidrule(lr){6-7} \cmidrule(lr){8-11}
                       & {\shortstack{Fact.\\Time (ms)}}
                       & {\shortstack{Solve\\Time (ms)}}
                       & {\shortstack{Total\\Time (ms)}}
                       & {\shortstack{Iter. \\ No.}}
                       & {\shortstack{Total\\Time  (ms)}}
                       & {\shortstack{Iter. \\ No.}}
                       & {\shortstack{Setup \\Time (ms)}}
                       & {\shortstack{Solve\\Time (ms)}}
                       & {\shortstack{Total\\Time (ms)}}
                       & {\shortstack{Iter. \\ No.}}
                       \\
    \midrule
    parabolic\_fem    & 21    & 237  & 527   & 40  & 467    & 923   & 52    & 16    & 68    & 10 \\
    \rowcolor{lightgray}
    ecology1          & 34    & 107  & 162   & 48  & 1146   & 1846  & 46    & 200   & 246   & 24 \\
    ecology2          & 34    & 106  & 163   & 49  & 1402   & 2181  & 74    & 22    & 96    & 11 \\
    \rowcolor{lightgray!40}
    apache2           & 49    & 94   & 177   & 25  & 723    & 1141  & 118   & 29    & 147   & 11 \\
    G3\_circuit       & 59    & 138  & 227   & 37  & 1033   & 1019  & 98    & 33    & 131   & 11 \\
    uniform Poisson   & 819   & 1780 & 2937  & 28  & 5174   & 256   & 1106  & 162   & 1268  & 9  \\
    aniso Poisson     & 442   & 324  & 941   & 10  & 8551   & 431   & 361   & 159   & 520   & 11 \\
    contrast Poisson  & 545   & 4851 & 5625  & 127 & 12544  & 638   & 516   & 194   & 710   & 12 \\
    spe16m            & 587   & 2028 & 2865  & 48  & 15437  & 694   & 440   & 210   & 650   & 13 \\
    GAP-road          & 481   & 2985 & 3607  & 106 & 213456 & \color{red}{10000} & 438 & 3311   & 3749  & 32 \\
    \rowcolor{lightgray}
    com-LiveJournal   & 26354 & 3698 & 35224 & 27  & 3517   & 95    & \multicolumn{4}{c}{\color{red}{Out of Memory}}  \\
    \rowcolor{lightgray}
    europe\_osm       & 1040  & 6041 & 7546  & 104 & 444754 & \color{red}{10000} & 873 & 10557 & 11430 & 28 \\
    delaunay\_n24     & 465   & 1420 & 2052  & 46  & 107135 & 4555  & 336   & 503   & 839   & 13 \\
    venturiLevel3     & 131   & 373  & 552   & 54  & 14756  & 4391  & 114   & 53    & 167   & 11 \\
    \rowcolor{lightgray}
    belgium\_osm      & 39    & 86   & 144   & 50  & 4201   & 5432  & 53    & 807   & 860   & 28 \\
    \bottomrule
  \end{tabular}
\end{table*}
%--------------------------------------------------------------------
\subsection{GPU Results}
\label{sec:gpu-results}

\begin{figure}[t]
  \centering
  \includegraphics[height=0.75\linewidth,width=0.75\linewidth]{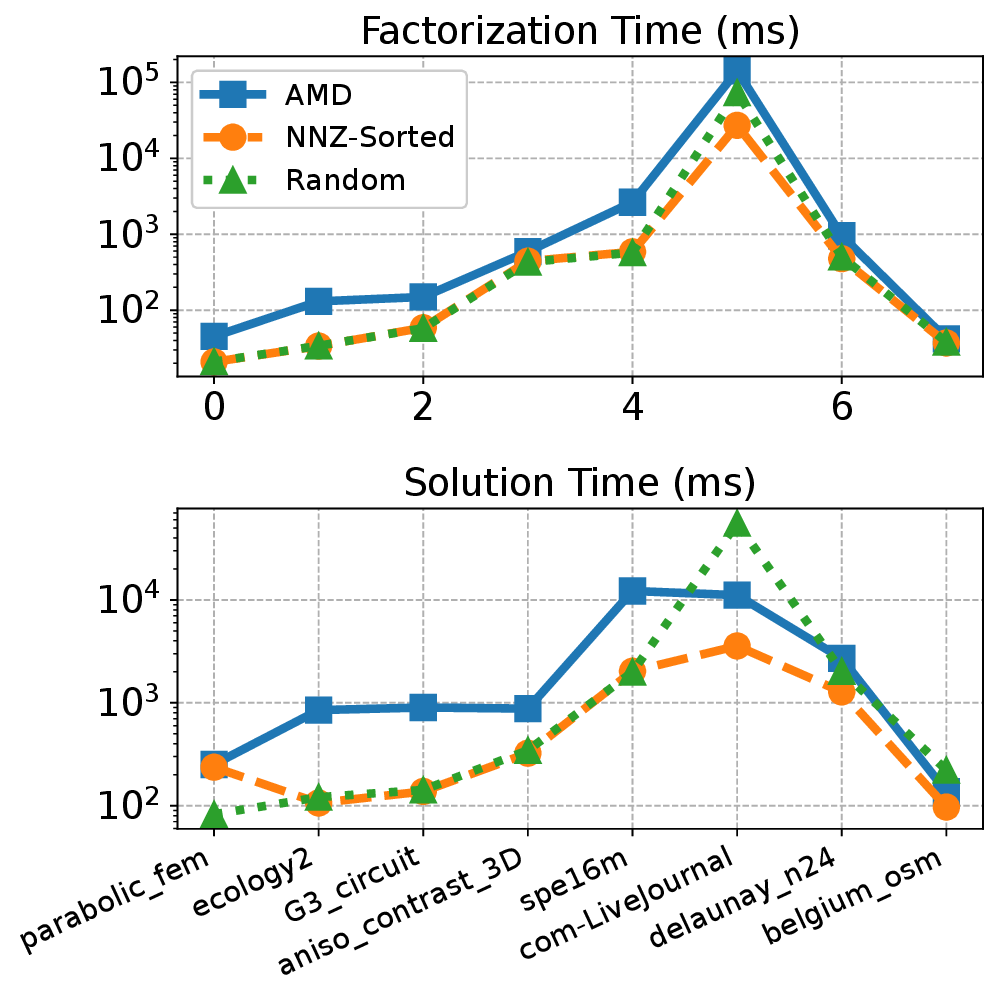}
  \caption{GPU performance under three orderings.}
\label{fig:gpu}
\end{figure}

% \begin{figure}[t]
%   \centering  
%   \includegraphics[height=1.2\linewidth,width=0.8\linewidth]{fig/gpu_depth.eps}

%   \caption{Comparison of three ordering algorithms in terms of critical-path length (equivalently, elimination-tree depth during the factorization phase). Top: exact Cholesky factorization. Middle: our approximate Cholesky factorization on GPU. Bottom: triangular solve using the approximate Cholesky factors.}
% \label{fig:depth}
% \end{figure}  

\paragraph{Impact of Ordering on GPU Performance}
We repeat the ordering study on the GPU with AMD, NNZ-sorted, and random. Although AMD is a strong CPU default, \Cref{fig:gpu} shows that \textbf{NNZ-sorted ordering} consistently outperforms it on the GPU.

Ordering affects both fill and dependency structure~\cite{liu1990role, liu1989reorder}. AMD is a fill-reducing heuristic~\cite{amd-order}; in our additional structural measurements, it produced deeper post-factorization elimination trees for the sampled \texttt{ParAC} factors and longer triangular-solve critical paths than NNZ-sorted or random ordering. These measurements indicate less available parallelism under AMD and help explain the GPU timing reversal.

The A100 also offers substantially higher peak memory bandwidth: NERSC reports 1,555\,GB/s for the 40\,GB A100 and 204.8\,GB/s per EPYC 7763 socket, a ratio of about $7.6\times$~\cite{nersc_perlmutter}. This likely benefits orderings that expose more concurrency, but does not by itself explain the reversal. We therefore recommend \textbf{NNZ-sorted} for GPU workloads. Random ordering is competitive on some matrices but less consistent overall.

\paragraph{Comparison with Baselines}
We compare the wall-clock performance of \texttt{ParAC} against two baselines: incomplete Cholesky with zero fill-in (IC(0)) from \texttt{cuSPARSE}, and the AMG solver in the \texttt{AmgX} library. \Cref{experiment:gpu} reports the results.

\textbf{Performance vs. \texttt{cuSPARSE} IC(0):}
\texttt{cuSPARSE} does not provide the classical elimination-based threshold ICT used in our CPU comparison, so our vendor-library GPU baseline is its IC(0) implementation. Unlike IC(0), \texttt{ParAC}'s fill pattern is not known \textit{a priori}; its GPU implementation therefore uses dynamic dependency tracking.

Among the 13 cases where both methods converge, \texttt{ParAC} improves total execution time on 11. The two exceptions are \texttt{parabolic\_fem}, where \texttt{ParAC} is about 13\% slower, and \texttt{com-LiveJournal}, a relatively dense graph on which the better preconditioning quality of \texttt{ParAC} does not offset its higher factorization cost. \texttt{ParAC} reduces the iteration count on every benchmark. It additionally converges on \texttt{GAP-road} and \texttt{europe\_osm}, where IC(0) fails to reach the $10^{-6}$ tolerance within $10{,}000$ iterations; \texttt{ParAC} converges on both in about $100$ iterations.

\textbf{Performance vs. \texttt{AmgX}} (through the \texttt{amgx\_capi.c} interface):
%
%\footnote{\url{https://github.com/NVIDIA/AMGX/blob/main/examples/amgx_capi.c}}:
%
Among the 14 cases completed by \texttt{AmgX}, its reported setup time exceeds \texttt{ParAC}'s reported factorization time in seven; the relative construction cost is therefore problem dependent. Its multilevel hierarchy exhausts the available GPU memory on \texttt{com-LiveJournal}, whereas \texttt{ParAC} completes. \texttt{ParAC} also outperforms \texttt{AmgX} on several other instances (gray rows). On the ill-conditioned \texttt{ecology1} matrix, for example, \texttt{AmgX} needs $24$ iterations and $200$\,ms of solve time. That is nearly an order of magnitude more solve time than on the closely related \texttt{ecology2} ($11$ iterations, $22$\,ms), and it is the reason \texttt{ParAC} wins on total time here. \texttt{ParAC} is also faster on \texttt{europe\_osm} and \texttt{belgium\_osm}; these cases show that \texttt{AmgX}'s advantage on the structured problems does not extend uniformly to the irregular networks in our suite.

\textbf{Effect of Matrix Density:}
The role of matrix density is clearest on \texttt{com-LiveJournal}, the densest matrix in our test suite (average degree $\approx 17$, see \Cref{table:inputs}).
On the CPU, both \texttt{hypre} and \texttt{ichol} are expensive here. \texttt{hypre} spends 236 seconds in setup, consistent with the cost of constructing a multilevel hierarchy for this dense, irregular graph, while \texttt{ichol} spends 194 seconds in factorization. \texttt{ParAC} completes the full solve in 35 seconds, roughly $6\times$ faster than \texttt{ichol} and $7\times$ faster than \texttt{hypre}.
On the GPU, the multilevel hierarchy of \texttt{AmgX} exhausts device memory outright (OOM). \texttt{cuSPARSE} IC(0) avoids this with zero fill-in and finishes in 3.5 seconds, ahead of the 35-second total for \texttt{ParAC}. It pays for this with $3.5\times$ more iterations (95 vs.\ 27), which reflects the trade-off between factorization cost and preconditioning quality on dense graphs.
Thus, \texttt{ParAC} avoids \texttt{AmgX}'s OOM while requiring fewer iterations than IC(0).

%\Cref{tab:gpu} compares the solution time and iteration counts among \texttt{ParAC}, \texttt{cuSPARSE}'s \texttt{ichol}, and \texttt{AmgX}. We see that \texttt{ParAC} outperforms \texttt{ichol} on most of the problems. Note that \texttt{cuSPARSE} uses a level-based strategy with zero fill-ins, which is different from MATLAB's threshold-based strategy for incomplete factorization. Algorithms with zero fill-ins tend to be faster in construction, but the preconditioning quality is worse. Hence, the number of iterations is significantly higher than the CPU case. For some problems, the iterative solver does not even converge.

%It is interesting to see that the AMG preconditioner, \texttt{AmgX}, performs the best for mesh-based problems, but it is inferior to \texttt{ParAC} on graph problems. It even runs out of memory on com-LiveJournal. GPU implementations of AMG preconditioners are less effective for problems out of the scientific domains.

%This matrix has a high density, which limits inter-block parallelism. Therefore, it is difficult for a GPU to take advantage of the bandwidth and the latency-hiding mechanism. 

%--------------------------------------------------------------------
\subsection{Robustness on Heterogeneous SDD Systems}
\label{sec:chimera}

To test robustness beyond the benchmark suite, we compare CPU \texttt{ParAC} with \texttt{hypre} and GPU \texttt{ParAC} with \texttt{AmgX} on 200 chimera systems from the SDDM2023 benchmark~\cite{gao2023robust}: 50 base graphs, each represented by weighted and unweighted variants, with and without boundary removal.
The heterogeneous construction recursively combines diverse graph families, including expanders, grids, trees, and random graphs.
The systems come from base graphs with $n = 100{,}000$ vertices; the boundary variants contain $97{,}872$ unknowns after boundary removal.
Within each platform comparison, the solvers use the same ordering and right-hand side; for boundary SDDM cases, \texttt{ParAC} uses the equivalent ground-node representation described above. CPU runs use AMD and GPU runs use NNZ-sorted ordering. The Chimera study uses a separate fixed \texttt{AmgX} configuration, also documented in the code repository. GPU \texttt{ParAC} used a fixed grid of 1,024 one-warp thread blocks and a $4\times$ edge pool; seven cases that did not complete with this setting were rerun with a $6\times$ pool.
All comparisons target a relative residual of $10^{-8}$. Because the solvers report residuals differently, we use a common acceptance threshold of $10^{-7}$ and require that a run neither crashes nor reaches the iteration cap.
\Cref{tab:chimera} summarizes the results.

\begin{table}[h]
  \centering
  \caption{Convergence on the 200 SDDM2023 chimera systems under the uniform acceptance criterion described in the text.}
  \label{tab:chimera}
  \small
  \begin{tabular}{lcc}
    \toprule
    Solver & Platform & Failures / 200 \\
    \midrule
    \texttt{hypre} & CPU & 15 \\
    \texttt{ParAC} & CPU & \textbf{0} \\
    \midrule
    \texttt{AmgX}  & GPU & 19 \\
    \texttt{ParAC} & GPU & \textbf{0} \\
    \bottomrule
  \end{tabular}
\end{table}

\texttt{hypre} fails on 15 systems: 10 weighted and five unweighted. These include crashes on all four variants of seed \texttt{s35}; the remaining runs either reach the iteration cap or terminate above the acceptance threshold.
\texttt{AmgX} fails on 19 full, no-boundary Laplacians, 18 weighted and one unweighted.
\texttt{ParAC} converges on all 200 systems in both platform studies. These results show greater robustness for \texttt{ParAC} than for both AMG baselines tested, without attributing every failure to a particular coarsening mechanism.

%--------------------------------------------------------------------

\section{Conclusions}
We introduced \texttt{ParAC}, a parallel execution framework for randomized incomplete Cholesky (IC) factorization. The numerical kernel is the AC randomized clique sampler of Gao, Kyng, and Spielman~\cite{gao2023robust}; the advance is an execution model based on online readiness rather than a precomputed separator hierarchy. Replacing static symbolic dependency estimation with \textbf{dynamic dependency tracking} lets \texttt{ParAC} resolve AC's stochastic sparsity pattern at runtime and exploit parallelism that cannot be scheduled statically in advance.

Our evaluation on the NERSC Perlmutter supercomputer spans 15 matrices from scientific and network domains. \texttt{ParAC} scales to 32 threads on AMD EPYC 7763 CPUs and to the fine-grained parallelism of NVIDIA A100 GPUs. Against deterministic IC, end-to-end speedups reach \textbf{57.3$\times$} on the CPU and \textbf{52.2$\times$} on the GPU, measured on cases where the baseline itself converges. On the irregular graph Laplacians tested, \texttt{ParAC} is up to \textbf{7.2$\times$} faster than AMG on the CPU and \textbf{6.0$\times$} faster on the GPU.

These results also delimit where \texttt{ParAC} belongs. For the structured PDE discretizations tested here, AMG generally gives lower total time. \texttt{ParAC} can be preferable on irregular graph Laplacians and SDDM systems, particularly when device memory cannot hold a multilevel hierarchy or when classical IC fails to converge on an irregular sparsity pattern. These situations arise in our experiments: \texttt{AmgX} exhausted GPU memory on \texttt{com-LiveJournal}, \texttt{cuSPARSE} IC(0) reached the iteration cap on two road networks, and the AMG solvers failed on 15 and 19 of the 200 chimera systems.

Several questions remain open. A mathematical characterization of the trade-off between fill reduction and structural parallelism would put the ordering guidance of \Cref{sec:gpu-results} on firmer ground. One promising direction draws on maximal independent set (MIS) algorithms: choose the elimination order dynamically as sampled fill changes, rather than fixing the permutation in advance~\cite{luby}. Supporting mixed-sign SDD matrices directly, rather than through the standard constant-factor reduction to a Laplacian~\cite{gremban1996combinatorial}, is another natural next step. The mechanism is not specific to Cholesky factorization either, and may extend to other randomized sparse primitives whose structure is revealed only at runtime.

A further direction is to use \texttt{ParAC} as a parallel smoother \emph{within} algebraic multigrid rather than as a standalone preconditioner. Recent work shows that incomplete Cholesky smoothing is effective inside a multigrid V-cycle and tolerates relatively low precision~\cite{vacek2025mixedmg}. On graph Laplacians, \texttt{ParAC} is breakdown-free by construction and parallelizes well, which makes it a natural candidate for that role on irregular problems. A smoother may also need far less accuracy than the spectral equivalence a preconditioner requires. How much sampling can be dropped under that weaker requirement is an open question.

\section*{Acknowledgment}
This work was supported in part by the U.S. Department of Energy, Office of Science, Office of Advanced Scientific Computing Research's Applied Mathematics Competitive Portfolios program under Contract No.\ DE-AC02-05CH11231.
The authors acknowledge support from Sparsitute, a Mathematical Multifaceted Integrated Capability Center funded by the U.S. Department of Energy, Office of Science, Advanced Scientific Computing Research.
We used resources of the National Energy Research Scientific Computing Center (NERSC), a Department of Energy Office of Science User Facility, under NERSC award ASCR-ERCAP-33069.
T.L.\ is supported by the National Science Foundation Graduate Research Fellowship Program under Grant No.\ 2146752.
H.L.\ was supported in part by U.S.\ National Science Foundation grant NSF-DMS 2412403.
Any opinions, findings, and conclusions or recommendations expressed in this material are those of the authors and do not necessarily reflect the views of the National Science Foundation.
The authors thank Shang Zhang, Steven Rennich, and Sergey Klevtsov of NVIDIA for their helpful insights.
AI-assisted tools, specifically OpenAI Codex and Anthropic Claude~\cite{openai_codex,anthropic_claude}, were used to edit wording throughout the manuscript and to check internal consistency and reference metadata. The authors reviewed and verified all suggestions; the research contributions, algorithm design, implementation, experiments, and reported results are their own.

%--------------------------------------------------------------------

\balance
\bibliographystyle{IEEEtran}
\bibliography{cite_doi}

@incollection{RugeStuben1987,
  author    = {John W. Ruge and Klaus St{\"u}ben},
  title     = {Algebraic Multigrid},
  booktitle = {Multigrid Methods},
  editor    = {Stephen F. McCormick},
  series    = {Frontiers in Applied Mathematics},
  volume    = {3},
  pages     = {73--130},
  publisher = {SIAM},
  address   = {Philadelphia, PA},
  year      = {1987},
  doi = {10.1137/1.9781611971057.ch4},
  url = {https://doi.org/10.1137/1.9781611971057.ch4}
}

@article{HensonYang2002,
  author  = {Van Emden Henson and Ulrike Meier Yang},
  title   = {{BoomerAMG}: A Parallel Algebraic Multigrid Solver and Preconditioner},
  journal = {Applied Numerical Mathematics},
  volume  = {41},
  number  = {1},
  pages   = {155--177},
  year    = {2002},
  doi     = {10.1016/S0168-9274(01)00115-5},
  url     = {https://doi.org/10.1016/S0168-9274(01)00115-5}
}

@inproceedings{AnztRibizelFlegarChowDongarra2019,
  author    = {Hartwig Anzt and Tobias Ribizel and Goran Flegar and Edmond Chow and Jack Dongarra},
  title     = {{ParILUT} - A Parallel Threshold {ILU} for {GPU}s},
  booktitle = {Proceedings of the 2019 IEEE International Parallel and Distributed Processing Symposium (IPDPS)},
  pages     = {231--241},
  year      = {2019},
  publisher = {IEEE},
  doi       = {10.1109/IPDPS.2019.00033},
  url       = {https://doi.org/10.1109/IPDPS.2019.00033}
}

@article{XuLiOseiKuffuor2025,
  author  = {Xu, Tianshi and Li, Ruipeng and Osei-Kuffuor, Daniel},
  title   = {A Two-Level {GPU}-Accelerated Incomplete {LU} Preconditioner for General Sparse Linear Systems},
  journal = {The International Journal of High Performance Computing Applications},
  volume  = {39},
  number  = {3},
  pages   = {424--442},
  year    = {2025},
  doi     = {10.1177/10943420251319334},
  url     = {https://doi.org/10.1177/10943420251319334}
}

@inproceedings{HysomPothen1999,
  author    = {David Hysom and Alex Pothen},
  title     = {Efficient Parallel Computation of {ILU}(k) Preconditioners},
  booktitle = {Proceedings of the 1999 ACM/IEEE Conference on Supercomputing},
  series    = {SC '99},
  pages     = {29},
  numpages  = {19},
  year      = {1999},
  publisher = {ACM},
  doi       = {10.1145/331532.331561},
  url       = {https://doi.org/10.1145/331532.331561}
}

@article{HysomPothen2001,
  author  = {David Hysom and Alex Pothen},
  title   = {A Scalable Parallel Algorithm for Incomplete Factor Preconditioning},
  journal = {SIAM Journal on Scientific Computing},
  volume  = {22},
  number  = {6},
  pages   = {2194--2215},
  year    = {2001},
  doi     = {10.1137/S1064827500376193},
  url     = {https://doi.org/10.1137/S1064827500376193}
}

@article{Haase1998,
  author  = {G. Haase},
  title   = {Parallel Incomplete {Cholesky} Preconditioners Based on the Non-Overlapping Data Distribution},
  journal = {Parallel Computing},
  volume  = {24},
  number  = {11},
  pages   = {1685--1703},
  year    = {1998},
  doi     = {10.1016/S0167-8191(98)00046-5},
  url     = {https://doi.org/10.1016/S0167-8191(98)00046-5}
}

@article{DoiWashio1999,
  author  = {Shun Doi and Takumi Washio},
  title   = {Ordering Strategies and Related Techniques to Overcome the Trade-Off Between Parallelism and Convergence in Incomplete Factorizations},
  journal = {Parallel Computing},
  volume  = {25},
  number  = {13--14},
  pages   = {1995--2014},
  year    = {1999},
  doi     = {10.1016/S0167-8191(99)00064-2},
  url     = {https://doi.org/10.1016/S0167-8191(99)00064-2}
}

@article{MagoluVanDerVorst2001FGCS,
  author  = {{Magolu monga Made}, M. and {van der Vorst}, H. A.},
  title   = {A Generalized Domain Decomposition Paradigm for Parallel Incomplete {LU} Factorization Preconditionings},
  journal = {Future Generation Computer Systems},
  volume  = {17},
  number  = {8},
  pages   = {925--932},
  year    = {2001},
  doi     = {10.1016/S0167-739X(01)00034-6},
  url     = {https://doi.org/10.1016/S0167-739X(01)00034-6}
}

@article{MagoluVanDerVorst2001ParCo,
  author  = {{Magolu monga Made}, M. and {van der Vorst}, H. A.},
  title   = {Parallel Incomplete Factorizations with Pseudo-Overlapped Subdomains},
  journal = {Parallel Computing},
  volume  = {27},
  number  = {8},
  pages   = {989--1008},
  year    = {2001},
  doi     = {10.1016/S0167-8191(01)00082-5},
  url     = {https://doi.org/10.1016/S0167-8191(01)00082-5}
}

@article{Watts1981,
  author  = {Watts, III, James W.},
  title   = {A Conjugate Gradient-Truncated Direct Method for the Iterative Solution of the Reservoir Simulation Pressure Equation},
  journal = {Society of Petroleum Engineers Journal},
  volume  = {21},
  number  = {3},
  pages   = {345--353},
  year    = {1981},
  doi     = {10.2118/8252-PA},
  url     = {https://doi.org/10.2118/8252-PA}
}

@article{Munksgaard1980,
  author  = {N. Munksgaard},
  title   = {Solving Sparse Symmetric Sets of Linear Equations by Preconditioned Conjugate Gradients},
  journal = {ACM Transactions on Mathematical Software},
  volume  = {6},
  number  = {2},
  pages   = {206--219},
  year    = {1980},
  doi     = {10.1145/355887.355893},
  url     = {https://doi.org/10.1145/355887.355893}
}

@article{JonesPlassmann1995,
  author  = {Mark T. Jones and Paul E. Plassmann},
  title   = {An Improved Incomplete {Cholesky} Factorization},
  journal = {ACM Transactions on Mathematical Software},
  volume  = {21},
  number  = {1},
  pages   = {5--17},
  year    = {1995},
  doi     = {10.1145/200979.200981},
  url     = {https://doi.org/10.1145/200979.200981}
}

@article{manteuffel1980incomplete,
  title={An incomplete factorization technique for positive definite linear systems},
  author={Manteuffel, Thomas A},
  journal={Mathematics of computation},
  volume={34},
  number={150},
  pages={473--497},
  year={1980},
  url={https://doi.org/10.1090/S0025-5718-1980-0559197-0},
  doi = {10.1090/S0025-5718-1980-0559197-0}
}

@phdthesis{gremban1996combinatorial,
  author  = {Gremban, Keith D.},
  title   = {Combinatorial Preconditioners for Sparse, Symmetric, Diagonally Dominant Linear Systems},
  school  = {Carnegie Mellon University},
  address = {Pittsburgh, PA},
  month   = oct,
  year    = {1996},
  note    = {{CMU} CS Tech. Rep. CMU-CS-96-123},
  url     = {https://www.cs.cmu.edu/~glmiller/Publications/b2hd-GrembanPHD.html}
}

@manual{cusparse,
  title = {{cuSPARSE} Library},
  author = {{NVIDIA Corporation}},
  year = {2024},
  url = {https://docs.nvidia.com/cuda/cusparse/},
  note = {Accessed: Apr. 1, 2026}
}

@misc{nersc_perlmutter,
  author       = {{National Energy Research Scientific Computing Center}},
  title        = {Perlmutter Architecture},
  howpublished = {NERSC Documentation},
  url          = {https://docs.nersc.gov/systems/perlmutter/architecture/},
  note         = {Accessed: Aug. 28, 2026}
}

@misc{openai_codex,
  author = {{OpenAI}},
  title  = {Codex},
  year   = {2026},
  url    = {https://developers.openai.com/codex/},
  note   = {Accessed: Aug. 28, 2026}
}

@misc{anthropic_claude,
  author = {{Anthropic}},
  title  = {Claude},
  year   = {2026},
  url    = {https://www.anthropic.com/claude},
  note   = {Accessed: Aug. 28, 2026}
}

@inproceedings{sachdeva2023simple,
  title={A simple and efficient parallel {Laplacian} solver},
  author={Sachdeva, Sushant and Zhao, Yibin},
  booktitle={Proceedings of the 35th ACM Symposium on Parallelism in Algorithms and Architectures},
  pages={315--325},
  year={2023},
  doi = {10.1145/3558481.3591101},
  url = {https://doi.org/10.1145/3558481.3591101},
}

@inproceedings{kyng2016approximate,
  title={Approximate {Gaussian} elimination for {Laplacian}s-fast, sparse, and simple},
  author={Kyng, Rasmus and Sachdeva, Sushant},
  booktitle={2016 IEEE 57th Annual Symposium on Foundations of Computer Science (FOCS)},
  pages={573--582},
  year={2016},
  organization={IEEE},
  doi = {10.1109/FOCS.2016.68},
  url = {https://doi.org/10.1109/FOCS.2016.68},
}

@ARTICLE{rosetarjan78,
    AUTHOR = {Rose, D.~J. and Tarjan, R.~E.},
    TITLE = {Algorithmic Aspects of Vertex Elimination on Directed Graphs},
    JOURNAL = {SIAM Journal on Applied Mathematics},
    VOLUME = {34},
    NUMBER = {1},
    PAGES = {176--197},
    YEAR = {1978},
    DOI = {10.1137/0134014},
    URL = {https://doi.org/10.1137/0134014}
}

@book{george1981computer,
  title={Computer Solution of Large Sparse Positive Definite Systems},
  author={George, Alan and Liu, Joseph W. H.},
  year={1981},
  publisher={Prentice Hall Professional Technical Reference}
}

@inproceedings{intel_wavefront,
author = {Venkat, Anand and Mohammadi, Mahdi Soltan and Park, Jongsoo and Rong, Hongbo and Barik, Rajkishore and Strout, Michelle Mills and Hall, Mary},
title = {Automating wavefront parallelization for sparse matrix computations},
year = {2016},
isbn = {9781467388153},
publisher = {IEEE Press},
booktitle = {Proceedings of the International Conference for High Performance Computing, Networking, Storage and Analysis},
articleno = {41},
numpages = {12},
location = {Salt Lake City, Utah},
series = {SC '16},
  doi = {10.1109/SC.2016.40},
  url = {https://doi.org/10.1109/SC.2016.40},
}

@article{chow2015fine,
  title={Fine-grained parallel incomplete {LU} factorization},
  author={Chow, Edmond and Patel, Aftab},
  journal={SIAM journal on Scientific Computing},
  volume={37},
  number={2},
  pages={C169--C193},
  year={2015},
  publisher={SIAM},
  doi = {10.1137/140968896},
  url = {https://doi.org/10.1137/140968896},
}

@techreport{naumov2012parallelLU,
  author      = {Naumov, Maxim},
  title       = {Parallel Incomplete-{LU} and {Cholesky} Factorization in the Preconditioned Iterative Methods on the {GPU}},
  institution = {NVIDIA},
  number      = {NVR-2012-003},
  month       = may,
  year        = {2012},
  url         = {https://research.nvidia.com/sites/default/files/pubs/2012-05_Incomplete-LU-and-Cholesky/nvr-2012-003.pdf}
}

@article{gao2023robust,
  title   = {{AC}($k$): Robust Solution of {Laplacian} Equations by Randomized Approximate {Cholesky} Factorization},
  author  = {Gao, Yuan and Kyng, Rasmus and Spielman, Daniel A.},
  journal = {SIAM Journal on Scientific Computing},
  volume  = {48},
  number  = {3},
  pages   = {A1643--A1684},
  year    = {2026},
  doi     = {10.1137/24M1673577},
  url     = {https://doi.org/10.1137/24M1673577}
}

@article{naumov2015amgx,
  title={{AmgX}: A library for {GPU} accelerated algebraic multigrid and preconditioned iterative methods},
  author={Naumov, Maxim and Arsaev, Marat and Castonguay, Patrice and Cohen, Jonathan and Demouth, Julien and Eaton, Joe and Layton, Simon and Markovskiy, Nikolay and Reguly, Istv{\'a}n and Sakharnykh, Nikolai and others},
  journal={SIAM Journal on Scientific Computing},
  volume={37},
  number={5},
  pages={S602--S626},
  year={2015},
  publisher={SIAM},
  URL = {https://doi.org/10.1137/140980260},
  doi = {10.1137/140980260}
}

@article{suitesparse,
author = {Davis, Timothy A. and Hu, Yifan},
title = {The university of {Florida} sparse matrix collection},
year = {2011},
issue_date = {November 2011},
publisher = {Association for Computing Machinery},
address = {New York, NY, USA},
volume = {38},
number = {1},
issn = {0098-3500},
url = {https://doi.org/10.1145/2049662.2049663},
doi = {10.1145/2049662.2049663},
journal = {ACM Trans. Math. Softw.},
month = dec,
articleno = {1},
numpages = {25}
}

@article{spe,
    author = {Christie, M. A. and Blunt, M. J.},
    title = {Tenth {SPE} Comparative Solution Project: A Comparison of Upscaling Techniques},
    journal = {SPE Reservoir Evaluation \& Engineering},
    volume = {4},
    number = {04},
    pages = {308--317},
    year = {2001},
    month = {08},
    issn = {1094-6470},
    doi = {10.2118/72469-PA},
    url = {https://doi.org/10.2118/72469-PA},
    eprint = {https://onepetro.org/REE/article-pdf/4/04/308/2586053/spe-72469-pa.pdf},
}

@article{liu1989reorder,
  title={Reordering sparse matrices for parallel elimination},
  author={Liu, Joseph W. H.},
  journal={Parallel Computing},
  volume={11},
  number={1},
  pages={73--91},
  year={1989},
  publisher={Elsevier},
  doi = {10.1016/0167-8191(89)90064-1},
  url = {https://doi.org/10.1016/0167-8191(89)90064-1},
}

@article{liu1990role,
  title={The role of elimination trees in sparse factorization},
  author={Liu, Joseph W. H.},
  journal={SIAM journal on matrix analysis and applications},
  volume={11},
  number={1},
  pages={134--172},
  year={1990},
  publisher={SIAM},
  doi = {10.1137/0611010},
  url = {https://doi.org/10.1137/0611010},
}

@article{rchol,
author = {Chen, Chao and Liang, Tianyu and Biros, George},
title = {{RCHOL}: Randomized {Cholesky} Factorization for Solving {SDD} Linear Systems},
journal = {SIAM Journal on Scientific Computing},
volume = {43},
number = {6},
pages = {C411--C438},
year = {2021},
doi = {10.1137/20M1380624},

URL = { 
    
        https://doi.org/10.1137/20M1380624
    
    

},
eprint = { 
    
        https://doi.org/10.1137/20M1380624
    
    

}
}

@misc{hypre,
  key   =       {hypre},
  title =       {{\sl hypre}: High Performance Preconditioners},
  note =        {\url{https://llnl.gov/casc/hypre}, \url{https://github.com/hypre-space/hypre}}
  }

@inproceedings{falgout2002hypre,
  title={{hypre}: A library of high performance preconditioners},
  author={Falgout, Robert D and Yang, Ulrike Meier},
  booktitle={International Conference on computational science},
  pages={632--641},
  year={2002},
  organization={Springer},
  url={https://doi.org/10.1007/3-540-47789-6_66},
  doi = {10.1007/3-540-47789-6_66}
}

@article{Davis2016DirectSurvey,
  title={A survey of direct methods for sparse linear systems},
  author={Timothy A. Davis and Sivasankaran Rajamanickam and Wissam M. Sid-Lakhdar},
  journal={Acta Numerica},
  year={2016},
  volume={25},
  pages={383--566},
  doi={10.1017/S0962492916000076},
  url={https://doi.org/10.1017/S0962492916000076}
}

@article{JMLR:learn-lap,
  author  = {Pierre Humbert and Batiste Le Bars and Laurent Oudre and Argyris Kalogeratos and Nicolas Vayatis},
  title   = {Learning {Laplacian} Matrix from Graph Signals with Sparse Spectral Representation},
  journal = {Journal of Machine Learning Research},
  year    = {2021},
  volume  = {22},
  number  = {195},
  pages   = {1--47},
  url     = {http://jmlr.org/papers/v22/19-944.html}
}

@inproceedings{learn-graph,
author = {Ando, Rie Kubota and Zhang, Tong},
title = {Learning on graph with {Laplacian} regularization},
year = {2006},
publisher = {MIT Press},
address = {Cambridge, MA, USA},
booktitle = {Proceedings of the 20th International Conference on Neural Information Processing Systems},
pages = {25--32},
numpages = {8},
location = {Canada},
series = {NIPS'06},
  doi = {10.7551/mitpress/7503.003.0009},
  url = {https://doi.org/10.7551/mitpress/7503.003.0009}
}

@article{elliptic-near-linear,
author = {Boman, Erik G. and Hendrickson, Bruce and Vavasis, Stephen},
title = {Solving Elliptic Finite Element Systems in Near-Linear Time with Support Preconditioners},
journal = {SIAM Journal on Numerical Analysis},
volume = {46},
number = {6},
pages = {3264--3284},
year = {2008},
doi = {10.1137/040611781},

URL = { 
    
        https://doi.org/10.1137/040611781
    
    

},
eprint = { 
    
        https://doi.org/10.1137/040611781
    
    

}
}

@inproceedings{spectral-partition,
author = {Guattery, Stephen and Miller, Gary L.},
title = {On the performance of spectral graph partitioning methods},
year = {1995},
isbn = {0898713498},
publisher = {Society for Industrial and Applied Mathematics},
address = {USA},
booktitle = {Proceedings of the Sixth Annual ACM-SIAM Symposium on Discrete Algorithms},
pages = {233--242},
numpages = {10},
location = {San Francisco, California, USA},
series = {SODA '95}
}

@book{saad-iterative,
author = {Saad, Yousef},
title = {Iterative Methods for Sparse Linear Systems},
publisher = {Society for Industrial and Applied Mathematics},
year = {2003},
doi = {10.1137/1.9780898718003},
address = {},
edition   = {Second},
URL = {https://epubs.siam.org/doi/abs/10.1137/1.9780898718003},
eprint = {https://epubs.siam.org/doi/pdf/10.1137/1.9780898718003}
}

@article{incomplete-chol,
title = {The incomplete {Cholesky}—conjugate gradient method for the iterative solution of systems of linear equations},
journal = {Journal of Computational Physics},
volume = {26},
number = {1},
pages = {43--65},
year = {1978},
issn = {0021-9991},
doi = {10.1016/0021-9991(78)90098-0},
url = {https://www.sciencedirect.com/science/article/pii/0021999178900980},
author = {David S Kershaw}
}

@inproceedings{luby,
author = {Luby, M},
title = {A simple parallel algorithm for the maximal independent set problem},
year = {1985},
isbn = {0897911512},
publisher = {Association for Computing Machinery},
address = {New York, NY, USA},
url = {https://doi.org/10.1145/22145.22146},
doi = {10.1145/22145.22146},
booktitle = {Proceedings of the Seventeenth Annual ACM Symposium on Theory of Computing},
pages = {1--10},
numpages = {10},
location = {Providence, Rhode Island, USA},
series = {STOC '85}
}

@article{superlu-overview,
author = {Li, Xiaoye S.},
title = {An overview of {SuperLU}: Algorithms, implementation, and user interface},
year = {2005},
issue_date = {September 2005},
publisher = {Association for Computing Machinery},
address = {New York, NY, USA},
volume = {31},
number = {3},
issn = {0098-3500},
url = {https://doi.org/10.1145/1089014.1089017},
doi = {10.1145/1089014.1089017},
journal = {ACM Trans. Math. Softw.},
month = sep,
pages = {302--325},
numpages = {24}
}

@article{gpu_sparse_chol,
title = {Accelerating sparse {Cholesky} factorization on {GPU}s},
journal = {Parallel Computing},
volume = {59},
pages = {140--150},
year = {2016},
note = {Theory and Practice of Irregular Applications},
issn = {0167-8191},
doi = {10.1016/j.parco.2016.06.004},
url = {https://www.sciencedirect.com/science/article/pii/S016781911630059X},
author = {Steven C. Rennich and Darko Stosic and Timothy A. Davis}
}

@article{amd-order,
author = {Amestoy, Patrick R. and Davis, Timothy A. and Duff, Iain S.},
title = {Algorithm 837: {AMD}, an approximate minimum degree ordering algorithm},
year = {2004},
issue_date = {September 2004},
publisher = {Association for Computing Machinery},
address = {New York, NY, USA},
volume = {30},
number = {3},
issn = {0098-3500},
url = {https://doi.org/10.1145/1024074.1024081},
doi = {10.1145/1024074.1024081},
journal = {ACM Trans. Math. Softw.},
month = sep,
pages = {381--388},
numpages = {8}
}

@inproceedings{baumann2024,
author = {Baumann, Yves and Kyng, Rasmus},
title = {A Framework for Parallelizing Approximate {Gaussian} Elimination},
year = {2024},
isbn = {9798400704161},
publisher = {Association for Computing Machinery},
address = {New York, NY, USA},
url = {https://doi.org/10.1145/3626183.3659987},
doi = {10.1145/3626183.3659987},
booktitle = {Proceedings of the 36th ACM Symposium on Parallelism in Algorithms and Architectures},
pages = {195--206},
numpages = {12},
location = {Nantes, France},
series = {SPAA '24}
}

@article{coarsening,
  title={Graph Coarsening: From Scientific Computing to Machine Learning},
  author={Jie Chen and Yousef Saad and Zechen Zhang},
  journal={SeMA Journal: Bulletin of the Spanish Society of Applied Mathematics},
  volume={79},
  number={1},
  pages={187--223},
  year={2022},
  doi = {10.1007/s40324-021-00282-x},
  url = {https://doi.org/10.1007/s40324-021-00282-x},
}

@book{davis2006direct,
  author    = {Timothy A. Davis},
  title     = {Direct Methods for Sparse Linear Systems},
  publisher = {SIAM},
  address   = {Philadelphia, PA},
  year      = {2006},
  doi       = {10.1137/1.9780898718881},
  isbn      = {9780898716139},
  url       = {https://doi.org/10.1137/1.9780898718881}
}

@article{anzt2018parilut,
  author  = {Hartwig Anzt and Edmond Chow and Jack J. Dongarra},
  title   = {{ParILUT}---A New Parallel Threshold {ILU} Factorization},
  journal = {SIAM Journal on Scientific Computing},
  volume  = {40},
  number  = {4},
  pages   = {C503--C519},
  year    = {2018},
  doi     = {10.1137/16M1079506},
  url     = {https://doi.org/10.1137/16M1079506}
}

@inproceedings{luo2025structilu,
  author    = {Hao Luo and Qianchao Zhu and Xiaochen Hao and Chunxi Lei and Chengdi Ma and Chenchen Zhang and Yun Liang and Chao Yang},
  title     = {{StructILU}: Dependency-Preserving Incomplete {LU} with Hierarchical Parallelism for Structured Grid {PDE}s on {GPU}s},
  booktitle = {Proceedings of the 39th ACM International Conference on Supercomputing},
  series    = {ICS '25},
  pages     = {119--134},
  year      = {2025},
  doi       = {10.1145/3721145.3725745},
  url       = {https://doi.org/10.1145/3721145.3725745}
}

@article{benzi2002preconditioning,
  author  = {Michele Benzi},
  title   = {Preconditioning Techniques for Large Linear Systems: A Survey},
  journal = {Journal of Computational Physics},
  volume  = {182},
  number  = {2},
  pages   = {418--477},
  year    = {2002},
  doi     = {10.1006/jcph.2002.7176},
  url     = {https://doi.org/10.1006/jcph.2002.7176},
}

@article{vacek2025mixedmg,
  author        = {Vacek, Petr and Anzt, Hartwig and Carson, Erin and Kohl, Nils and R\"{u}de, Ulrich and Tsai, Yu-Hsiang},
  title         = {Mixed precision multigrid with smoothing based on incomplete {Cholesky} factorization},
  journal       = {arXiv preprint arXiv:2511.04566},
  year          = {2025},
  url           = {https://doi.org/10.48550/arXiv.2511.04566},
  doi = {10.48550/arXiv.2511.04566}
}

@inproceedings{powerrchol,
  author    = {Zhiqiang Liu and Wenjian Yu},
  title     = {{PowerRChol}: Efficient Power Grid Analysis Based on Fast Randomized {Cholesky} Factorization},
  booktitle = {Proceedings of the 61st ACM/IEEE Design Automation Conference},
  series    = {DAC '24},
  articleno = {190},
  pages     = {1--6},
  year      = {2024},
  doi       = {10.1145/3649329.3657308},
  url       = {https://doi.org/10.1145/3649329.3657308}
}

@misc{laplaciansjl,
  author       = {Daniel A. Spielman},
  title        = {{Laplacians.jl}: Algorithms Inspired by Graph {Laplacians}},
  howpublished = {\url{https://github.com/danspielman/Laplacians.jl}},
  note         = {{J}ulia package},
  year         = {2020}
}

%\appendix
%\input{appendix_v4}

\end{document}